\documentclass{aa}  

\usepackage{silence}
\usepackage{graphicx}
\usepackage{txfonts}
\usepackage{array}
\usepackage{siunitx}
\usepackage{pdflscape}
\usepackage{multicol}
\usepackage{booktabs}

\vbadness=\maxdimen

\usepackage{natbib}
\usepackage[breaklinks]{hyperref}
\bibpunct{(}{)}{;}{a}{}{,}
\hypersetup{
  colorlinks,
  citecolor=blue,
  linkcolor=blue,
  urlcolor=blue,
}

\renewcommand{\arraystretch}{1.4}
\newcolumntype{M}[1]{>{\raggedright \arraybackslash}m{#1}}
\newcolumntype{C}[1]{>{\centering \arraybackslash}m{#1}}
\newcolumntype{R}[1]{>{\raggedright\arraybackslash}p{#1}}

\newcolumntype{f}[1]{>{\footnotesize \raggedright \arraybackslash}m{#1}}

\newcolumntype{B}{>{\bfseries \boldmath}l}

\graphicspath{{./images/}}

\DeclareSIUnit\angstrom{\text {Å}}

\begin{document}

\title{Super-Eddington accretion and early-stage feedback in Ton S180}

\subtitle{}

\author{
Pierpaolo Condò\inst{1,2,3,4} \corrauth{pierpaolo.condo@inaf.it}
\and
Giacomo Venturi \inst{5} \email{giacomo.venturi1@sns.it}
\and
Eleonora Parlanti \inst{5} \email{eleonora.parlanti@sns.it}
\and
Marco Berton \inst{2} \email{Marco.Berton@eso.org}
\and
Enrico Congiu \inst{2} \email{econgiu@eso.org}
\and
Alessia Tortosa \inst{3} \email{alessia.tortosa@inaf.it}
\and
Stefano Carniani \inst{5} \email{stefano.carniani@sns.it}
\and
Francesco Tombesi \inst{1,3,4} \email{francesco.tombesi@roma2.infn.it}
\and
Tommaso Zana \inst{6,3,7} \email{tommaso.zana@uniroma1.it}
\and
Claudio Ricci \inst{8,9} \email{claudio.ricci.astro@gmail.com}
\and
Ezequiel Treister \inst{10} \email{treister@gmail.com}
\and
Luis C. Ho \inst{11,12} \email{lho.pku@gmail.com}
\and
Enrico Piconcelli \inst{3} \email{enrico.piconcelli@inaf.it}
\and
Miguel Coloma Puga \inst{2,13,14} \email{miguel.colomapuga@inaf.it}
\and
Emilia Järvelä \inst{15} \email{astrojarvela@gmail.com}
\and
Luca Crepaldi \inst{16, 17} \email{luca.crepaldi@unipd.it}
\and
Amelia Vietri \inst{16,18} \email{amelia.vietri@unipd.it}
\and
Benedetta Dalla Barba \inst{17} \email{benedetta.dallabarba@inaf.it}
}

\institute{
Dipartimento di Fisica, Università di Roma Tor Vergata, Via della Ricerca Scientifica, 1, Roma 00133, Italy
\and
European Southern Observatory (ESO), Alonso de Córdova 3107, Casilla 19, Santiago, Chile
\and
INAF, Osservatorio Astronomico di Roma, Via Frascati 33, I-00040 Monte Porzio Catone, Italy
\and
INFN, Rome Tor Vergata, Via della Ricerca Scientifica 1, 00133 Rome, Italy
\and
Scuola Normale Superiore, Piazza dei Cavalieri 7, I-56126 Pisa, Italy
\and
Dipartimento di Fisica, Sapienza, Università di Roma, Piazzale Aldo Moro 5, IT-00185 Roma, Italy
\and
INFN, Sezione di Roma I, Piazzale Aldo Moro 2, IT-00185 Roma, Italy
\and
Department of Astronomy, University of Geneva, ch. d'Ecogia 16, 1290, Versoix, Switzerland
\and
Instituto de Estudios Astrof\'isicos, Facultad de Ingenier\'ia y Ciencias, Universidad Diego Portales, Av. Ej\'ercito Libertador 441, Santiago, Chile
\and
Instituto de Alta Investigaci{\'{o}}n, Universidad de Tarapac{\'{a}}, Casilla 7D, Arica, Chile
\and
Kavli Institute for Astronomy and Astrophysics, Peking University, Beijing 100871, People's Republic of China
\and
Department of Astronomy, School of Physics, Peking University, Beijing 100871, People's Republic of China
\and
Dipartimento di Fisica, Università degli Studi di Torino, Via Pietro Giuria 1, 10125 (Torino), Italy
\and
INAF – Osservatorio Astrofisico di Torino, Via Osservatorio 20, I-10025 Pino Torinese, Italy
\and
Department of Physics \& Astronomy, Texas Tech University, Lubbock TX 79409-1051, USA
\and
Dipartimento di Fisica e Astronomia ``G. Galilei'', Università di Padova, Vicolo dell’Osservatorio 3, 35122, Padova, Italy
\and
Osservatorio Astronomico di Brera, Istituto Nazionale di Astrofisica (INAF), Via E. Bianchi 46, Merate (LC) 23807, Italy
\and
INAF – Astronomical Observatory of Padova, Vicolo dell'Osservatorio 5, 35122, Padova, Italy
}

\date{Received 10 May 2026 / Accepted 1 September 2026}

  \abstract{
  Narrow-line Seyfert 1 (NLSy1) galaxies are key laboratories for studying rapid supermassive black hole (SMBH) growth and active galactic nucleus (AGN) feedback at high accretion rates. We investigate the nearby NLSy1 Ton S180 with VLT-MUSE optical integral field spectroscopy to connect its nuclear accretion properties with the spatially resolved ionized gas and host-galaxy kinematics. We modeled the unresolved nuclear spectrum and applied a custom point spread function subtraction to recover the host-galaxy emission on kiloparsec scales. The nuclear spectrum requires a complex permitted-line decomposition and a blueshifted [\ion{O}{iii}] outflow component. Single-epoch estimators and the stellar velocity dispersion imply black hole masses in the range of $\log(M_{\rm BH}/M_\odot)=6.5$--$7.7$. Combined with the observed luminosity, this implies a dimensionless mass accretion rate of $\dot{M} / \dot{M}_{\rm{Edd}}=4.1$--$980$, confirming the extreme accretion regime. The host galaxy shows a circumnuclear ring, an inner elongated structure consistent with a bar, and rotation-dominated gas and stellar kinematics. Simple inflow models do not reproduce the observed velocity field significantly better. We detect a resolved ionized outflow extending about 2 kpc west of the nucleus, with mildly blueshifted velocities (with a maximum of $v_{\rm max} \sim 340$ km s$^{-1}$). Its mass outflow rate is only $\sim 4.2 \times 10^{-4}\,M_\odot\,\mathrm{yr}^{-1}$, whereas the unresolved nuclear outflow reaches $v_{\rm max} \sim 1140$ km s$^{-1}$ and $\dot{M}_{\rm out} > 1.5\,M_\odot\,\mathrm{yr}^{-1}$. This contrast may reflect either weak ionized coupling from nuclear to galactic scales or different episodes of AGN activity over time.
  These results show that Ton S180 is undergoing rapid SMBH growth, while the observed ionized outflow remains confined to the inner few kiloparsecs and is weak on host-galaxy scales.
  }
  \keywords{galaxies: active, galaxies: nuclei, galaxies: Seyfert, galaxies: individual: Ton S180, galaxies: individual: HE0054-2239}

\maketitle
\nolinenumbers
\titlerunning
\authorrunning

\section{Introduction}
\label{sec:intro}
Supermassive black holes (SMBHs), with masses spanning $10^6$--$10^{10} \, M_\odot$, reside at the centers of massive galaxies. When actively accreting interstellar material, they power active galactic nuclei (AGNs), with luminosities of $L > 10^{42}$ erg s$^{-1}$ across the electromagnetic spectrum. In the local Universe, scaling relations between SMBH mass ($M_{\mathrm{BH}}$) and host-galaxy properties, such as the $M_{\mathrm{BH}}$--$\sigma_\star$ and $M_{\mathrm{BH}}$--$M_{\mathrm{bulge}}$ relations, imply coevolution regulated by nongravitational processes \citep{ferrarese_fundamental_2000, reines_relations_2015}. Energetic outflows likely mediate this interaction \citep{silk_quasars_1998, king_black_2003}, spanning multiple scales and gas phases, from highly ionized, sub-parsec ultrafast outflows (UFOs) detected as blueshifted X-ray absorption features near the innermost stable circular orbit \citep{tombesi_evidence_2010, xrism_pdsNature_2025}, to ionized winds traced by optical and UV lines \citep{fiore_agn_2017}, and neutral atomic and molecular outflows on kiloparsec scales \citep[e.g.,][]{cicone_largely_2018}.

The standard paradigm for sub-Eddington accretion, where the accretion luminosity remains below the Eddington limit\footnote{The Eddington luminosity, $L_{\mathrm{Edd}} = 1.26 \times 10^{38} (M_{\rm BH}/M_\odot)$ erg s$^{-1}$, represents the luminosity at which radiation pressure balances gravity.}, is the geometrically thin, optically thick Shakura-Sunyaev disk \citep[SSD;][]{shakura_black_1973}, in which gas pressure ($P_g$) dominates radiation pressure ($P_{\mathrm{rad}}$). The disk maintains a thin aspect ratio ($H/r \leq 0.1$, where $H$ is the scale height and $r$ is the radius) and emits a multi-temperature blackbody spectrum peaking in the optical--UV.
The strength of the accretion flow is parametrized by the normalized accretion rate, $\dot{m}$, defined as
\begin{equation}
\label{eq:mdot}
\dot{m} = \dot{M} / \dot{M}_{\rm{Edd}} \, ,
\end{equation}
where $\dot{M}$ is the physical mass accretion rate through the disk and $\dot{M}_{\mathrm{Edd}} = L_{\mathrm{Edd}}/c^2$ is the Eddington accretion rate. 
The observed Eddington ratio is defined by\footnote{The definition of $\dot{m}$ varies in the literature: some works incorporate the radiative efficiency, $\eta$ \citep[e.g.,][]{netzer_physics_2013}, yielding $\dot{m}=\lambda_{\mathrm{Edd}}$, while others, including this work, do not \citep[e.g.,][]{yuan_hot_2014}.}
\begin{equation}
\label{eq:lamba_Edd}
\lambda_{\mathrm{Edd}} = L_{\mathrm{bol}}/L_{\mathrm{Edd}} = \eta \dot{m} \, ,
\end{equation}
where $\eta= L_{\mathrm{bol}} / (\dot{M}c^2)$ is the actual radiative efficiency; that is, the fraction of the rest-mass energy supply rate that is radiated away. The thin-disk solution is typically expected for $10^{-2} \lesssim \dot{m} \lesssim 1$; at lower accretion rates, advection-dominated solutions become relevant.
Under our adopted definition, at high accretion rates, roughly $\dot{m} \gtrsim 10$, inward advection becomes important, and the standard radiatively efficient SSD description becomes inadequate \citep{abramowicz_foundations_2013, czerny_slim_2019}. A widely adopted description is the geometrically thick slim-disk framework \citep{abramowicz_slim_1988}, subsequently refined by relativistic treatments \citep{sadowski_slim_2009}. In these solutions, large optical depths and short inflow times trap photons, which are advected inward rather than escaping. The radiative efficiency, $\eta$, therefore falls below the theoretical maximum, $\eta_{\rm max} = 6\%$--$42\%$, set by the innermost stable circular orbit and the black hole spin.
Because of this photon trapping, the bolometric luminosity ($L_{\mathrm{bol}}$) decouples from the true mass accretion rate, and $\lambda_{\mathrm{Edd}}$ saturates with only a logarithmic dependence on $\dot{m}$ \citep{marziani_superedd_2025}. Alternative radiatively efficient super-Eddington channels have also been proposed \citep{socrates_ultraluminous_2006}. Furthermore, radiation pressure and non-axisymmetric instabilities drive powerful outflows \citep{meier_structure_1982, ohsuga_supercritical_2005, ohsuga_global_2009, massonneau_how_2023}, a regime increasingly explored in numerical simulations within both idealized and cosmological environments \citep{Zana2026, lupi_sustained_2024}.
We term a system with $\dot{m} < 1$ subcritical, whereas supercritical denotes systems with $\dot{m} > 1$. Because the radiative efficiency, $\eta$, depends on the accretion rate, $\dot{m}$ and $\lambda_{\mathrm{Edd}}$ are not strictly interchangeable: a system may be supercritical in mass supply ($\dot{m} > 1$) while remaining only marginally sub-Eddington in luminosity ($\lambda_{\mathrm{Edd}} < 1$).

Narrow-line Seyfert 1 galaxies (NLSy1s) are key local laboratories for these accretion regimes. They are typically defined by (i) narrow broad permitted lines with full width at half maximum FWHM(H$\beta$) $<2000$ km s${}^{-1}$, (ii) weak [\ion{O}{iii}]~$\lambda 5007$ emission with [\ion{O}{iii}]/H$\beta <3$, and (iii) often prominent \ion{Fe}{ii} complexes with $R_{4570} \gtrsim 1$ \citep[e.g.,][]{osterbrock_spectra_1985}. NLSy1s consistently exhibit high accretion rates, $\dot{m}\sim 0.1$--100, and relatively low SMBH masses, $M_{\mathrm{BH}}\sim 10^{6}$--$10^{8}\,M_{\odot}$ \citep[e.g.,][]{boroson_emission-line_1992, berton_how_2025}.
Permitted emission-line profiles in NLSy1s typically show a narrow core (full width at half maximum [FWHM]~$\sim$~400--800~km~s$^{-1}$) associated with the narrow-line region (NLR), plus broader wings that often require at least two Gaussian components: an intermediate-width one (FWHM~$\sim$~900--1400~km~s$^{-1}$) and a very broad one (FWHM~$>$~2000~km~s$^{-1}$).
For H$\beta$, \citet{paul_analysis_2022} found that the width of the intermediate component is consistent with the width of the Fe~\textsc{ii} emission, although no statistical correlation was found. The strong Fe~\textsc{ii} complexes, which are characteristic of NLSy1s, remain difficult to model. Standard photoionization does not reproduce their strength, suggesting that additional processes such as Ly$\alpha$ fluorescence or shock heating may contribute \citep[see, e.g.,][]{baldwin_origin_2004, cracco_spectroscopic_2016}.
X-ray observations further underscore the exotic nature of NLSy1s: steep spectra, flux variability exceeding a factor of 2 on hour-long timescales in the 0.3–10 keV band, and frequent detections of UFOs ($v \sim 0.1$--$0.3\,c$; \citealt{boller_rapid_1993, laurenti_ufo_2026}).
The energetics of these UFOs (kinetic power $\dot{E}_{\rm kin} \equiv \tfrac{1}{2}\dot{M}_{\rm out}v^2$, often with $\dot{E}_{\rm kin}/L_{\rm bol} \gtrsim 0.05$) can theoretically quench star formation by ejecting gas reservoirs or heating the circumgalactic medium \citep{tombesi_evidence_2012, king_powerful_2015}. In cosmological simulations, AGN feedback is required to reproduce the high-mass truncation of the galaxy luminosity function \citep{croton_many_2006, vogelsberger_model_2013}.

Recent \textit{James Webb} Space Telescope (JWST) discoveries further highlight the relevance of NLSy1s for high-redshift accreting black holes. The growing JWST census of low-luminosity broad-line AGNs at high redshifts includes systems resembling local NLSy1s, with relatively narrow broad lines, low black hole masses, and high, sometimes super-Eddington, accretion rates \citep{juodzbalis_direct_2025, berton_how_2025, maiolino_small_2024, tortosa_hyperion_2024}. Important differences remain: many sources appear overmassive relative to the local $M_{\mathrm{BH}}$--$M_\star$ relation and are unusually X-ray weak, especially among little red dots \citep{juodzbalis_direct_2025, maiolino_jwst_2025, kocevski_rise_2025, Tortosa2026}.
High signal-to-noise (S/N) JWST H$\alpha$ spectra also show that single Gaussians are inadequate, whereas Lorentzian and multi-Gaussian profiles often match or outperform exponential profiles, supporting a stratified broad-line region (BLR) interpretation \citep{scholtz_little_2026}. Local NLSy1s can therefore serve as high-resolution laboratories for accretion physics that likely also operated in the early Universe \citep{berton_how_2025, Tortosa2026}.
In parallel, host-galaxy dynamics remain central to the fueling problem: bars and resonance rings are plausible channels for inward gas transport, but linking kiloparsec-scale noncircular motions to instantaneous BH feeding is still observationally challenging.

In this work, we study the nuclear region and the spatially resolved ionized gas properties of Ton S180, a highly accreting NLSy1 in the local Universe, using VLT-MUSE optical integral field spectroscopy. Our three primary goals are to map the ionized gas kinematics and search for outflow signatures beyond the nucleus, characterize the spatially resolved excitation mechanisms and morphology of the host galaxy, and look for deviations from pure circular motions that could indicate gas inflows feeding the accretion.
Section~\ref{sec:tons180} summarizes the source properties, and Sect.~\ref{sec:data_red} describes the data reduction. The analysis is presented in Sect.~\ref{sec:data_analysis}, including the unresolved nuclear spectrum (Sect.~\ref{sec:agnspec}), the AGN subtraction (Sect.~\ref{sec:spatial_analysis}), the spatially resolved ionization and outflow analysis (Sects.~\ref{sec:bpt} and \ref{sec:outflow}), and the gas and stellar kinematics (Sect.~\ref{sec:kinematics_method}). We then present the physical results in Sect.~\ref{sec:results}, discuss their implications in Sect.~\ref{sec:discussion}, and summarize our conclusions in Sect.~\ref{sec:conclusions}. We assume a flat $\Lambda$ cold dark matter cosmology with $H_0 = 67.7$ km s$^{-1}$ Mpc$^{-1}$, $\Omega_\mathrm{m} = 0.311$, and $\Omega_\Lambda = 0.689$.

\begin{figure} 
  \centering
  \includegraphics{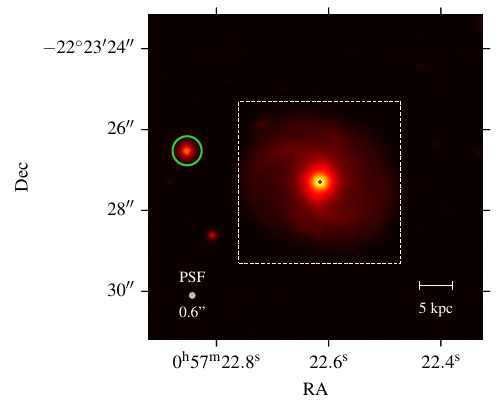}
  \caption{MUSE white-light image of Ton S180. The centroid of the bright AGN nucleus is marked with a cross. In the left corner we report the average point spread function (PSF) full width at half maximum (FWHM) of 0.6\,arcsec over the $4750-9350$~\AA\ spectral range. The white bar in the bottom left corner indicates a scale of 5 kpc at the redshift of the target. The dashed box marks the 100$\times$100 pixels sub-cube used for the galaxy analysis in this work. The field star marked with a green circle was used to construct the PSF model for the aperture correction of the nuclear spectrum (Appendix~\ref{appendix:psfmod}).}
  \label{fig:CubeImage}
\end{figure}

\begin{figure*}
  \centering
  \includegraphics[width=\linewidth]{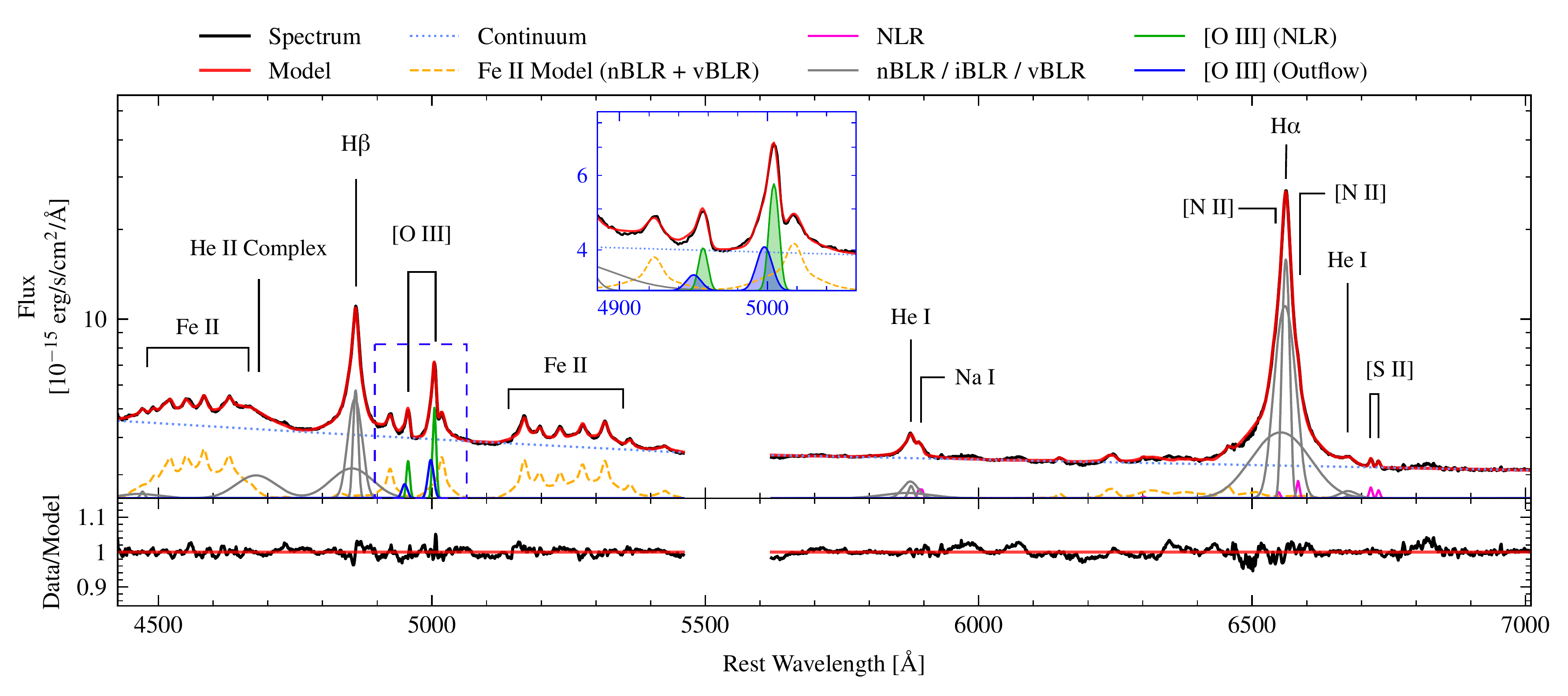}
  \caption{Unresolved nuclear spectrum of Ton S180 extracted from the four spaxels that sample the peak of the nuclear PSF. The best-fit of the active galaxy using the fiducial three-Gaussian (3G) model is overplotted. For the Balmer and He lines, the narrow (nBLR), intermediate (iBLR), and broad (vBLR) components are plotted separately in gray. The narrow line emission is displayed in pink. The \ion{Fe}{ii} emission is modeled with two Gaussians (narrow and intermediate), which are summed and shown as a dashed yellow line for clarity. The blueshifted [\ion{O}{iii}] lines and their blue wing, tracing an outflowing component of the narrow line region, are highlighted in green and blue, respectively. The inset shows a zoom-in of the [\ion{O}{iii}] line region. The bottom panel shows the ratio between the observed data and the model fit.
  }
  \label{fig:AGNSpec}
\end{figure*}

\section{Ton S180}
\label{sec:tons180}

Ton S180 (also known as HE~0054$-$2239) is a local ($z = 0.0622$) NLSy1 galaxy, located at right ascension $0^{\mathrm{h}}57^{\mathrm{m}}22.6^{\mathrm{s}}$ and declination $-22^\circ23'27''$ (J2000), with a bolometric luminosity $\log(L_{\mathrm{bol}}/\mathrm{erg\,s}^{-1}) \approx 45.2$ \citep{scharwachter_spatially_2017}. Its location near the South Galactic Pole makes local dust extinction negligible ($E(B-V)=0.0124$; \citealt{1998ApJ...500..525S}). In X-rays, Ton S180 is a bare NLSy1, lacking intrinsic absorption \citep{turner_chandra_2001}, and its most striking property is its extreme Eddington ratio.
\citet{kawaguchi_origin_2004} modeled its spectral energy distribution (SED) using a self-gravitating slim disk plus outer non-Keplerian structure, deriving an intrinsic $\dot{m} \approx 1000$ and $\log(M_{\mathrm{BH}}/M_\odot) \approx 6.8$.
Conversely, \citet{matzeu_first_2020} applied the \textsc{agnslim} model, tailored for super-Eddington accretion, to joint XMM-Newton and NuSTAR data, finding $\dot{m} \approx 5$ and $\log(M_{\mathrm{BH}}/M_\odot) \approx 7$.
Based on optical spectra, \citet{scharwachter_spatially_2017} report $\lambda_{\mathrm{Edd}}=1.7$ \footnote{We note that, while the previous estimates were, consistently with our notation, relative to the dimensionless mass accretion rate ($\dot{m}$), the latter ($\lambda_{\mathrm{Edd}}$) considers the Eddington ratio defined as the bolometric luminosity over the Eddington luminosity, which includes the efficiency factor (Eq.~\ref{eq:lamba_Edd}).}.
Despite the difference, all the studies agree about the very high accretion rate.
However, optical reverberation mapping (RM) monitoring campaigns are currently unavailable for Ton S180, hindering a direct, model-independent determination of the black hole mass. 
The lack of optical RM data introduces considerable uncertainties in the black hole mass estimation, which compound with the systematic inaccuracies of virial mass estimators, particularly for high-$\dot{m}$ systems \citep[e.g.,][]{dalla_bonta_sloan_2020}.
In the optical, Ton S180 shows classic NLSy1 features: strong \ion{Fe}{ii} emission (\ion{Fe}{ii}/H$\beta \approx 0.9$), H$\beta$ FWHM $\approx 1060$ km s${}^{-1}$ \citep{scharwachter_spatially_2017}. The [\ion{O}{iii}] line exhibits a blueward asymmetry, indicative of nuclear outflows \citep{paul_analysis_2022} and has a line width of [\ion{O}{iii}]~$\lambda5007$ FWHM $\approx 630$ km s${}^{-1}$ \citep{scharwachter_spatially_2017}.
The host galaxy is disk-dominated with spiral arms and a possible classical bulge \citep{mathur_supermassive_2012}. Its star formation rate, SFR $\sim 0.8\,M_\odot\,\text{yr}^{-1}$, is estimated from the extinction-corrected H$\alpha$ luminosity using the relation from \citet{kennicutt_star_1998}. Its specific rate, $\mathrm{sSFR} = \mathrm{SFR}/M_\star \sim 10^{-10.8}\,\text{yr}^{-1}$, is intermediate between the values typical of main-sequence and quiescent galaxies, placing it in the green valley \citep{scharwachter_spatially_2017}.
Previous integral field spectroscopy with WiFeS \citep{wifes_instrument_2007} hinted at large-scale rotation, but its $\sim$3 kpc resolution could not resolve sub-kiloparsec structures or detect extended nuclear outflows \citep{scharwachter_spatially_2017}. The X-ray spectrum shows a soft excess and a steep power-law tail, typical of highly accreting NLSy1s, best described by a two-corona scenario (warm optically thick and hot optically thin regions) linked to super-Eddington accretion and a slim disk \citep{matzeu_first_2020}. XMM-Newton data also reveal a highly ionized outflow at $v_{\mathrm{out}} \approx -0.2c$, consistent with the extreme accretion regime and relativistic outflows \citep{matzeu_first_2020}.

\section{Data reduction}
\label{sec:data_red}
We observed Ton S180 with the Multi Unit Spectroscopic Explorer (MUSE; \citealt{Bacon:2010a}) at the Very Large Telescope (VLT) of the European Southern Observatory (ESO). The observations used the wide field mode (WFM) with ground layer adaptive optics (GLAO), providing a field of view of approximately $1'\times1'$ with a spatial sampling of $0.2''$ per spaxel. For our combined dataset, the observations span the spectral range 4750--9350\,\AA\ and achieve a measured PSF FWHM varying between $0.5$--$0.8''$, depending on the wavelength, with an average of $\sim0.6''$ (i.e., $\sim0.75$ kpc). These data, obtained under ESO program 111.24SR.001 (PI G. Venturi) on June 28--29, 2023, consist of 18 exposures of 240\,s each, for a total on-source integration time of 4320\,s. The atmospheric seeing measured at the start of the exposures varied between $0.73''$ and $1.86''$, while the GLAO correction delivered the effective PSF width reported above, enabling sub-kiloparsec-resolution analysis.
We reduced the data using the ESO MUSE pipeline \citep{Weilbacher:2020a} within the \textsc{ESO Reflex} environment, which manages the Common Pipeline Library recipes \citep{Banse:2004a, ESOCPL2015}. The reduction process included bias subtraction, flat fielding, illumination correction, and wavelength and flux calibration. We also performed the geometric reconstruction of the data cube and the final combination of all exposures. Sky subtraction used background regions within the field of view free from source emission. The pipeline's automatic masking provided an accurate sky model, and the final cube showed no evidence of line over-subtraction. A white-light image of the final combined cube is shown in Fig.~\ref{fig:CubeImage}.
We refined the absolute astrometry by cross-matching two field stars (Fig.~\ref{fig:CubeImage}) against the \textit{Gaia} DR3 catalog \citep{GaiaDR3} and applied the resulting coordinate offset to the final cube.

\section{Data analysis}
\label{sec:data_analysis}

\subsection{Unresolved AGN spectral analysis}
\label{sec:agnspec}
\begin{table}
\tiny
\centering
\caption{Best-fit parameters for the prominent emission lines in the unresolved AGN spectrum of Ton S180 for the fiducial three-Gaussian model.}
\label{tab:AGNfit}
\setlength{\tabcolsep}{4pt}
\renewcommand{\arraystretch}{1.2}
\begin{tabular}{lccc}

\toprule
Line & Flux & FWHM & Velocity Offset \\
\midrule
\multicolumn{4}{c}{very broad line region (vBLR)} \\
H$\beta$ 4861 & $61 \pm 5$ & $5500 \pm 100$ & $-540 \pm 20$ \\
H$\alpha$ 6563 & $190 \pm 10$ &  &  \\
\ion{He}{i} 5877 & $16 \pm 1$ &  &  \\
\ion{He}{ii} $\lambda$4686 Complex & $37_{-3}^{+4}$ & $> 4000$ & $> 400$ \\

\midrule
\multicolumn{4}{c}{intermediate broad line region (iBLR)} \\
H$\beta$ 4861 & $79 \pm 6$ & $1800_{-40}^{+60}$ & $-136 \pm 8$ \\
H$\alpha$ 6563 & $310_{-30}^{+20}$ &  &  \\
\ion{He}{i} 4471 & $3.5 \pm 1$ &  &  \\
\ion{He}{i} 5877 & $16 \pm 1$ &  &  \\
\ion{He}{i} 6678 & $5.7_{-0.8}^{+0.9}$ &  &  \\

\midrule
\multicolumn{4}{c}{narrow broad line region (nBLR)} \\
H$\beta$ 4861 & $40_{-2}^{+3}$ & $630_{-20}^{+30}$ & $-52 \pm 3$ \\
H$\alpha$ 6563 & $210 \pm 10$ &  &  \\
\ion{He}{i} 4471 & $1.5_{-0.6}^{+0.7}$ &  &  \\
\ion{He}{i} 5877 & $3.5_{-0.4}^{+0.7}$ &  &  \\
\ion{He}{i} 6678 & $1.1_{-0.4}^{+0.5}$ &  &  \\

\midrule
\multicolumn{4}{c}{Narrow Line Region (NLR)} \\
\ion{Na}{i} 5890 & $1.4_{-0.3}^{+0.5}$ & $300_{-30}^{+50}$ & $0 \pm 10$ \\
\ion{Na}{i} 5896 & $1.5_{-0.3}^{+0.5}$ &  &  \\
{[\ion{O}{i}]} 6300 & $<0.4$ &  &  \\
{[\ion{N}{ii}]} 6583 & $5 \pm 2$ &  &  \\
{[\ion{S}{ii}]} 6716 & $1.6_{-0.3}^{+0.5}$ &  &  \\
{[\ion{S}{ii}]} 6731 & $1.2_{-0.3}^{+0.4}$ &  &  \\
{[\ion{O}{iii}]} 5007 (core) & $22 \pm 3$ & $380 \pm 20$ & $-170 \pm 30$ \\
{[\ion{O}{iii}]} 5007 (wing) & $16 \pm 3$ & $730_{-30}^{+50}$ & $-520_{-40}^{+30}$ \\

\midrule
\multicolumn{4}{c}{Iron Multiplets} \\
\ion{Fe}{ii} (broad) & --- & $> 3000$ & $40 \pm 50$ \\
\ion{Fe}{ii} (narrow) & --- & $680 \pm 30$ & $-25_{-8}^{+6}$ \\
\bottomrule
\end{tabular}
\tablefoot{
  The model overlaid on the data is shown in Figs.~\ref{fig:AGNSpec} and \ref{fig:modelcomp}, bottom panel. Fluxes are given in units of $10^{-15}$ erg s$^{-1}$ cm$^{-2}$; FWHM and velocity offsets are given in kilometers per second. Errors are at the 68\% confidence level.
  Velocity offset values are relative to the zero point of the kinematic modeling (see Sect.~\ref{sec:kinematics_method}).}
\end{table}
To analyze the nucleus of Ton S180, we extracted the integrated spectrum from the four brightest spaxels in the MUSE data cube, which enclose the centroid and peak of the nuclear PSF. An aperture correction based on the MUSE PSF model (see Appendix~\ref{appendix:psfmod}) was applied to recover the total nuclear flux. An aperture correction factor of $\sim4.8$ was adopted, since these spaxels encompass only 20\% of the total PSF flux. A linear decomposition of the MUSE nuclear spectrum using AGN and host SDSS eigenspectra \citep{yip_spectral_2004}, following the approach described by \citet{ilic_fantastic_2023}, returned a negligible host-galaxy contribution, consistent with the overwhelming AGN brightness.
The observed spectrum presents several notable features:
\begin{enumerate}
    \item Two Gaussian components are insufficient to model the H$\alpha$ and H$\beta$ profiles, which require at least an additional intermediate-width component.
    \item The [\ion{O}{iii}] $\lambda\lambda$4959, 5007 lines are systematically blue-shifted relative to other lines and present a blue wing indicating a distinct outflowing component.
    \item The \ion{Fe}{ii} emission is both kinematically complex (not well described by a single broad Gaussian) and cannot be reproduced by standard empirical Seyfert templates (e.g., \citealt{veron_templates_2004}), requiring updated atomic data such as those presented in \citet{ilic_fantastic_2023}.
\end{enumerate}
We therefore optimized several spectral models with $\chi^2$ statistics, using the Bayesian Information Criterion (BIC; \citealt{kass1995bayes}) to compare their complexity and fit quality. These models were built with our custom Python framework \texttt{prism}, described briefly in Appendix~\ref{appendix:prism}.
The kinematics of the BLR, NLR, \ion{Fe}{ii}, and [\ion{O}{iii}] components were allowed to vary independently.
Theoretical flux ratios for [\ion{N}{ii}], [\ion{O}{iii}] and [\ion{O}{i}] doublets were fixed at 1/3 \citep{osterbrock_astrophysics_2006}.
Instrumental broadening was accounted for by convolving the model profiles with the MUSE line spread function to get intrinsic line widths (see Appendix \ref{appendix:prism} for details). Throughout this paper, all reported values of FWHM and velocity dispersion refer to these intrinsic, deconvolved quantities. 
We tested both broad-band and H$\beta$-focused fits and found consistent parameters, although the broad-band fit leaves larger continuum residuals; unless stated otherwise, we adopt the broad-band solution as the most complete description.
Our fiducial model, hereafter referred to as Model 3G, uses three Gaussian components for the permitted lines named the very broad line region (vBLR), intermediate broad line region (iBLR), and narrower broad line region (nBLR), and two Gaussian components for \ion{Fe}{ii}. It reproduces the nuclear spectrum well (Fig.~\ref{fig:AGNSpec}; Table~\ref{tab:AGNfit}) and is statistically preferred over the simpler alternatives, with the quantitative BIC comparison summarized in Appendix~\ref{appendix:agnmodel}. The 4640--4700 \AA\ feature listed in Table~\ref{tab:AGNfit} as the ``\ion{He}{ii} $\lambda$4686 Complex'' is treated as an unresolved blend, likely including \ion{He}{ii}, Bowen fluorescence lines, and \ion{Fe}{ii}; fitting it as a single \ion{He}{ii} line would yield unphysical fluxes comparable to or larger than H$\beta$. The comparison with the two-Gaussian and Gaussian+Lorentzian models, together with additional tests on \ion{Fe}{ii} kinematics and the implications of the newly detected vBLR component for the black hole mass, is presented in Appendix~\ref{appendix:agnmodel}.
The best-fit parameters for the most prominent emission lines are listed in Table~\ref{tab:AGNfit}, based on the fiducial three-Gaussian model.
Velocity offsets are measured relative to the systemic redshift ($z=0.06219 \pm 0.00015$), determined from the zero point of the dynamical model of the host galaxy's 2D kinematics (see Sect.~\ref{sec:kinematics_method}). Hereafter, all velocities reported in this work, including in the maps, are relative to this systemic redshift. We preferred this approach to inferring the systemic velocity from an integrated spectrum since the host galaxy absorption features are not detectable in the nuclear spectrum and we cannot rely on narrow emission lines that may be outflowing or affected by other kinematic disturbances.
Errors are reported at the 68\% confidence level and are derived from $N=1000$ Monte Carlo resamplings of the spectrum, taking into account the MUSE error spectrum.
Instrumental broadening was taken into account using the MUSE line response which is approximately Gaussian with $\sigma=50$ km s$^{-1}$ at the H$\alpha$ wavelength (see Appendix \ref{appendix:prism} for implementation details).

\subsection{Unresolved AGN emission subtraction}
\label{sec:spatial_analysis}
To study the host galaxy within the central few kiloparsecs, we first removed the unresolved AGN emission, whose PSF wings dominate the inner field and outshine the extended line emission by several orders of magnitude. We adopted an empirical subtraction based on the observed nuclear spectrum, rescaled spaxel by spaxel to account for the wavelength-dependent PSF shape. This custom approach is well suited to the extreme nucleus-to-host contrast of Ton S180 and recovers the narrow line emission across the circumnuclear region.
The subtraction module was implemented within \texttt{prism}, linking empirical AGN removal and line fitting in the same framework. The full implementation, validation tests, and an example are given in Appendix~\ref{appendix:psfmod}. Several alternative PSF subtraction methods were tested but failed in this contrast regime. Our empirical subtraction yields a continuum-subtracted cube for spatially resolved emission-line analysis, while intentionally sacrificing stellar-continuum information in the innermost region to recover the narrow gaseous component robustly.

\begin{figure*} 
  \centering
  \includegraphics[width=\linewidth]{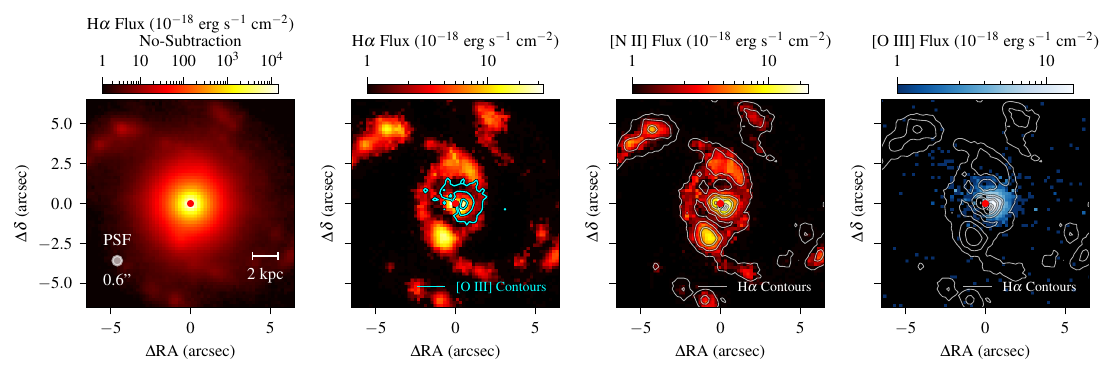}
  \caption{\textit{Left to right}: Original H$\alpha$ flux map, AGN-subtracted H$\alpha$ flux map, and AGN-subtracted [\ion{N}{ii}] and [\ion{O}{iii}] flux maps.
  The second panel has the contours of the AGN-subtracted [\ion{O}{iii}] flux map overlaid for reference, while the third and fourth panels show the contours of the AGN-subtracted H$\alpha$ flux map.
  The circle marks the position of the nuclear emission peak; the four central spaxels used to define the nuclear template are masked in the residual maps, as they are assumed to be entirely dominated by unresolved AGN emission.
  The original H$\alpha$ map is dominated by the AGN PSF, which obscures the underlying galaxy structure. After subtraction, the H$\alpha$ map reveals a circumnuclear ring that connects the spiral arms and an elongated east--west structure within the ring that could trace a bar. The [\ion{N}{ii}] map has a morphology overlapping with that of H$\alpha$. The [\ion{O}{iii}] map is more compact and shows a clear westward skew relative to the nucleus, likely tracing the NLR.
  }
  \label{fig:fluxmaps}
\end{figure*}

\subsection{Emission-line fitting of the AGN-subtracted cube}
\label{sec:emline_fitting}
To analyze the emission lines in the AGN-subtracted cube, we performed a spatially resolved spectral fit with \texttt{prism}, using the same model definitions adopted for the nuclear spectrum.
Emission lines were modeled as single Gaussians with tied kinematic parameters (FWHM and velocity offset) across H$\alpha$, H$\beta$, [\ion{O}{i}], [\ion{O}{iii}], [\ion{N}{ii}], and [\ion{S}{ii}]. An example of the subtraction and residual line fitting is shown in Appendix~\ref{appendix:psfmod}.
This approach provides a reliable description of the subtracted sub-cube, yielding spatially resolved maps of line fluxes and kinematics with an effective spatial resolution of about 0.6\arcsec\ in the emission-line windows used here (see Appendix~\ref{appendix:psfmod}). At the redshift of the target, this corresponds to approximately 0.8 kpc.
The average reduced $\tilde{\chi}^2={\chi}^2/\text{d.o.f.}$ of the fit is 1.42, with a standard deviation $\sigma_{\tilde{\chi}^2}=0.18$ across the field of view, indicating a good overall fit quality. Figure~\ref{fig:fluxmaps} shows the resulting flux maps for H$\alpha$, [\ion{N}{ii}], and [\ion{O}{iii}] compared to the original cube.
The H$\alpha$ flux map reveals extended spiral arms reaching about 10\,kpc from the nucleus and a prominent circumnuclear ring connecting these arms. We also identify an elongated east–west structure within the ring that potentially represents a nuclear bar. In contrast, the [\ion{O}{iii}] emission is significantly more compact than H$\alpha$ and shows a clear westward skew relative to the nucleus, which likely traces the NLR.

\subsection{Spatially resolved ionization diagnostics}
\label{sec:bpt}
\begin{figure} 
  \centering
  \includegraphics[width=\linewidth]{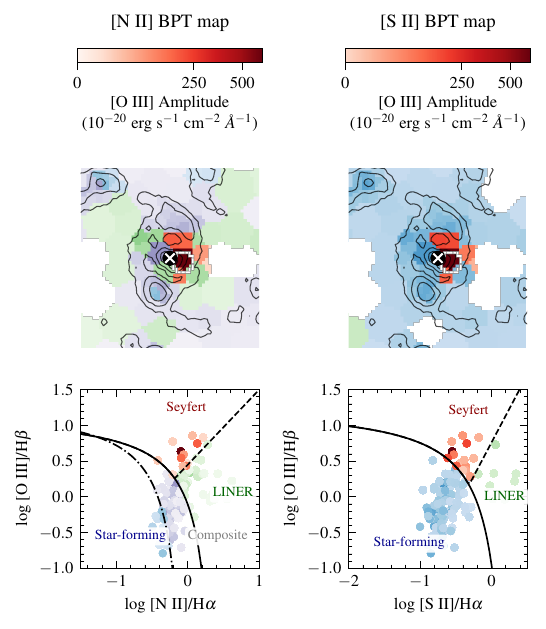}
  \caption{\textit{Bottom panels}: [\ion{O}{iii}] $\lambda5007$/H$\beta$ versus [\ion{N}{ii}] $\lambda6583$/H$\alpha$ (left) and [\ion{S}{ii}] $\lambda\lambda6716,6731$/H$\alpha$ (right) spatially resolved diagnostic diagrams for the AGN-subtracted data cube, considering the narrow components only (i.e., without the broad component observed in [\ion{O}{iii}]). In the [N II] diagram, the curves from \citet[][solid]{BPTKewley2001} and \citet[][dash-dotted]{BPTKauffmann2003}  mark the theoretical maximal and the empirical star-forming boundaries, respectively, while the \citet[][dashed]{BPTSchawinski2007} line separates Seyfert-like and LINER-like excitation in the AGN regime. In the [S II] diagram, the \citet[][solid]{BPTKewley2001} curve separates star-forming from AGN-like ionization, and the \citet[][dashed]{BPTKewley2006} line divides Seyfert-like and LINER-like excitation. Most bins fall in the composite or star-forming regions, while bins on the west side of the nucleus, shown in red, are AGN-dominated. \textit{Top panels}: Spatial distribution of the bins color-coded by their classification in the corresponding BPT diagram. The white cross marks the location of the unresolved nuclear emission, the gray contour indicates the outflow extraction region. The discrete colors identify the BPT classes, while the transparency of each bin scales with the [\ion{O}{iii}] peak amplitude, as traced by the color bars.}
  \label{fig:bpt}
\end{figure}

To characterize the ionization mechanisms across the host galaxy, we used [\ion{O}{iii}]/H$\beta$, [\ion{S}{ii}]/H$\alpha$ and [\ion{N}{ii}]/H$\alpha$ emission-line diagnostic diagrams \citep{baldwin_bpt_1981, veilleux_bpt_1987} based on the best-fit maps from the AGN-subtracted cube. Similar spatially resolved ionization analyses in nearby AGNs have been presented by \citet{venturi_magnum_2018, venturi_complex_2023} and \citet{finlez_luminosity_2025}.
Since the S/N of individual spaxels in the AGN-subtracted cube is relatively low, especially for H$\beta$ and [\ion{O}{iii}], the resulting diagnostic diagrams are noisy and do not reveal clear spatial trends on a spaxel-by-spaxel basis. We therefore performed adaptive Voronoi binning of the cube \citep{cappellari_voronoi_2003}, which ensures a minimum S/N per bin while maximizing spatial resolution. For the BPT analysis, we used the H$\beta$ signal and noise maps as input for the binning, since H$\beta$ is the limiting line entering both the [\ion{N}{ii}]- and [\ion{S}{ii}]-based diagnostic diagrams, and we required a minimum S/N of five per bin. 
We then computed [\ion{O}{iii}]/H$\beta$ against [\ion{N}{ii}]/H$\alpha$ and ([\ion{S}{ii}] $\lambda6716 +$ [\ion{S}{ii}] $\lambda6731$)/H$\alpha$ for each bin, requiring S/N $> 2$ for all lines involved. Here the S/N is measured as the ratio of the fit line peak to the RMS noise estimated in the 5075--5125\,\AA\ continuum window.
Figure~\ref{fig:bpt} shows the resulting diagnostic diagram and the spatial distribution of the bins. The spiral arms and circumnuclear ring are mostly classified as composite, likely reflecting a mix of stellar and AGN photoionization, while the west side of the nucleus, co-spatial with the [\ion{O}{iii}] peak, is AGN-dominated. This analysis guided our definition of the region where we searched for a resolved ionized outflow.

\subsection{Spatially resolved outflow identification}
\label{sec:outflow}
As discussed in Sect.~\ref{sec:agnspec}, the unresolved nuclear spectrum already shows a complex [\ion{O}{iii}] profile, with a strongly blueshifted wing and a narrower core that is itself offset from the systemic velocity. To test whether this outflow extends beyond the unresolved nucleus, we examined the [\ion{O}{iii}] kinematics in the AGN-subtracted cube.
\begin{figure} 
  \centering
  \includegraphics{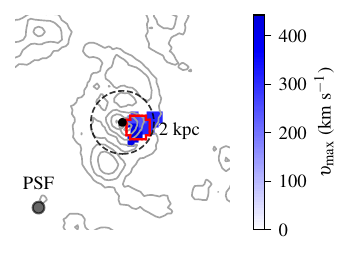}
  \includegraphics{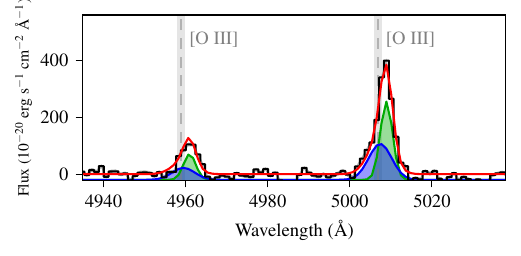}
  \caption{\textit{Top}: [\ion{O}{iii}] outflow maximum velocity map ($v_{\rm max} = |v_{\rm off}| + 2\sigma_{\rm out}$) derived from the second Gaussian component in the two-component fit of the binned cube, which we associate with the spatially resolved ionized outflow. Only bins where the second component is required with $\Delta \mathrm{BIC} > 10$ are shown. The black dot marks the location of the unresolved nuclear emission as traced by the spatial peak of the broad Balmer lines and the continuum. The red contour indicates the area where the integrated spectrum (bottom panel) was extracted to derive the average outflow properties, while the dashed black circle marks a projected galactocentric radius of 2\,kpc, which we adopted as the maximum outflow radius. \textit{Bottom}: Integrated spectrum of the outflow region (black) with the best-fit model (red) and its decomposition into core (green) and outflowing (blue) components for the [\ion{O}{iii}] emission. The vertical dashed lines indicate the [\ion{O}{iii}] transitions at the systemic velocity of the galaxy and its associated error as derived from the kinematic modeling zero point velocity (see Sect.~\ref{sec:kinematics_method}). A second broad component is required to reproduce the blueshifted wing of the [\ion{O}{iii}] line, confirming the presence of an extended ionized outflow.}
  \label{fig:outflow}
\end{figure}
We first fit all emission lines (H$\alpha$, H$\beta$, [\ion{O}{i}], [\ion{O}{iii}], [\ion{N}{ii}], and [\ion{S}{ii}]) with a single Gaussian component. We then added a second Gaussian only for the [\ion{O}{iii}] doublet, because no comparable wing is detected in the other lines, and used $\Delta$BIC to test whether the extra component was statistically required. In the initial spaxel-by-spaxel fits, the detections remained too patchy to define a convincing structure, so we repeated the analysis on the Voronoi-binned cube described above.

In the binned cube, we detected a spatially coherent region with a maximum extent of $2.0 \pm 0.5$~kpc where the second [\ion{O}{iii}] component is statistically required. Here $\Delta \mathrm{BIC} = \mathrm{BIC}_1 - \mathrm{BIC}_2$, where the one- and two-component models are compared in each bin, and both BIC and Akaike information criterion tests favor the additional component. We retained only bins with $\Delta \mathrm{BIC} > 10$, resulting in eight bins with a robust detection (Fig.~\ref{fig:outflow}, top panel). Velocity offsets, $v_{\rm off}$, were measured relative to the systemic redshift ($z = 0.06219 \pm 0.00015$) consistently with the nuclear outflow methodology. Across these eight bins, the outflow component line centroid offset ranges from $v_{\rm off} \approx +20$ to $+60$\,km\,s$^{-1}$ relative to systemic velocity, corresponding to a blueshift of about $-20$ to $-100$\,km\,s$^{-1}$ relative to the local rotating host-galaxy disk velocity ($v_{\rm rot} \approx +123$\,km\,s$^{-1}$ at this radius, Sect.~\ref{sec:kinematics_method}).

To derive representative outflow parameters, we extracted a spectrum from a circular aperture of radius $r=3$ spaxels (i.e., 0.75~kpc) centered on the resolved outflow region (Fig.~\ref{fig:outflow}). In this integrated spectrum, a two-component model is strongly preferred, with $\Delta \mathrm{BIC} > 260$. The best-fit outflow component has $\mathrm{FWHM} = 350 \pm 10$\,km\,s$^{-1}$ ($\sigma_{\rm out} = 149 \pm 4$\,km\,s$^{-1}$) and a centroid velocity offset of $v_{\rm off} = 38 \pm 6$\,km\,s$^{-1}$ relative to systemic. Relative to the local rotating host disk, the gas is distinctly blueshifted by $v_{\rm off} - v_{\rm rot} = -84 \pm 6$\,km\,s$^{-1}$.
Considering the host-galaxy inclination ($i \approx 37^\circ$; Sect.~\ref{sec:host}) and disk obscuration, dust extinction likely hides the counter-facing side of the flow, while projection of a tilted biconical wind on the western side naturally accounts for a mildly positive line-of-sight velocity centroid while remaining photoionized by the AGN cone (Sect.~\ref{sec:bpt}).
In any case, given the uncertainty on the host-galaxy systemic redshift ($\delta z = 0.00015$, corresponding to $\approx \pm 42$\,km\,s$^{-1}$), finding a redshifted centroid per se is not a statistically robust outcome, as it remains consistent with zero or a slight blueshift within $1\sigma$.
Adopting the outflow velocity definition of $v_{\rm max} = |v_{\rm off}| + 2\sigma_{\rm out}$, which assumes that $v_{\rm max}$ is the intrinsic outflow velocity and that the lower line-of-sight velocities described by the broad component spectral profile are due to projection effects \citep[see e.g.,][]{Rupke2013, Venturi2026}, this yields a representative maximum outflow velocity of $v_{\rm max} = 340 \pm 10$\,km\,s$^{-1}$.

\subsection{Gas and stellar kinematics of the host galaxy}
\label{sec:kinematics_method}
\begin{figure} 
  \centering
  \includegraphics{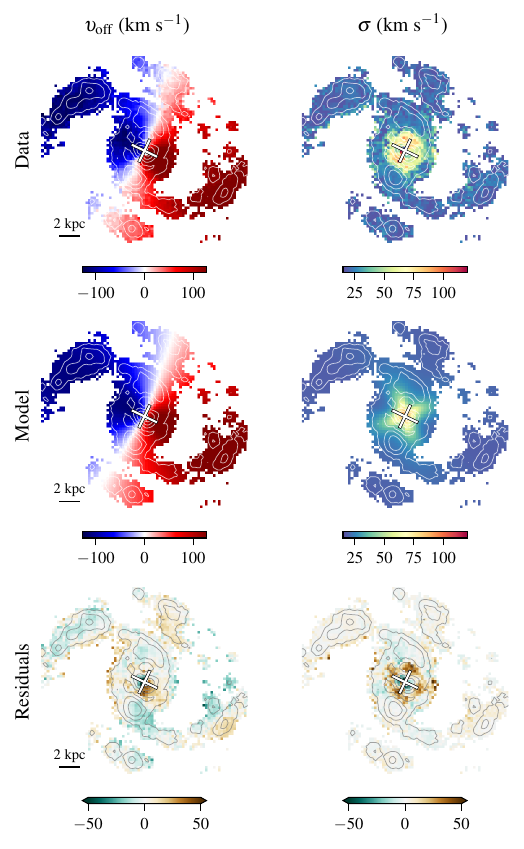}
  \caption{Maps of host galaxy kinematics derived from the AGN-subtracted data cube. \textit{Top row}: Ionized gas line-of-sight velocity (left) and velocity dispersion (right). \textit{Middle row}: Best-fit kinematic models assuming pure circular motion. \textit{Bottom row}: Residuals. The kinematic center is marked by the white cross, whose major axis is aligned to the kinematic position angle and minor axis length is scaled by $\cos(i)$ to reflect the galaxy inclination to the line of sight.
  The velocity field is reproduced well by circular rotation, while residuals reaching $\sim 50$ km s$^{-1}$ along the spiral-arm region suggest non-axisymmetric streaming motions.
  }
  \label{fig:kinematic_model}
\end{figure}
\begin{figure}
  \centering
  \includegraphics[width=\linewidth]{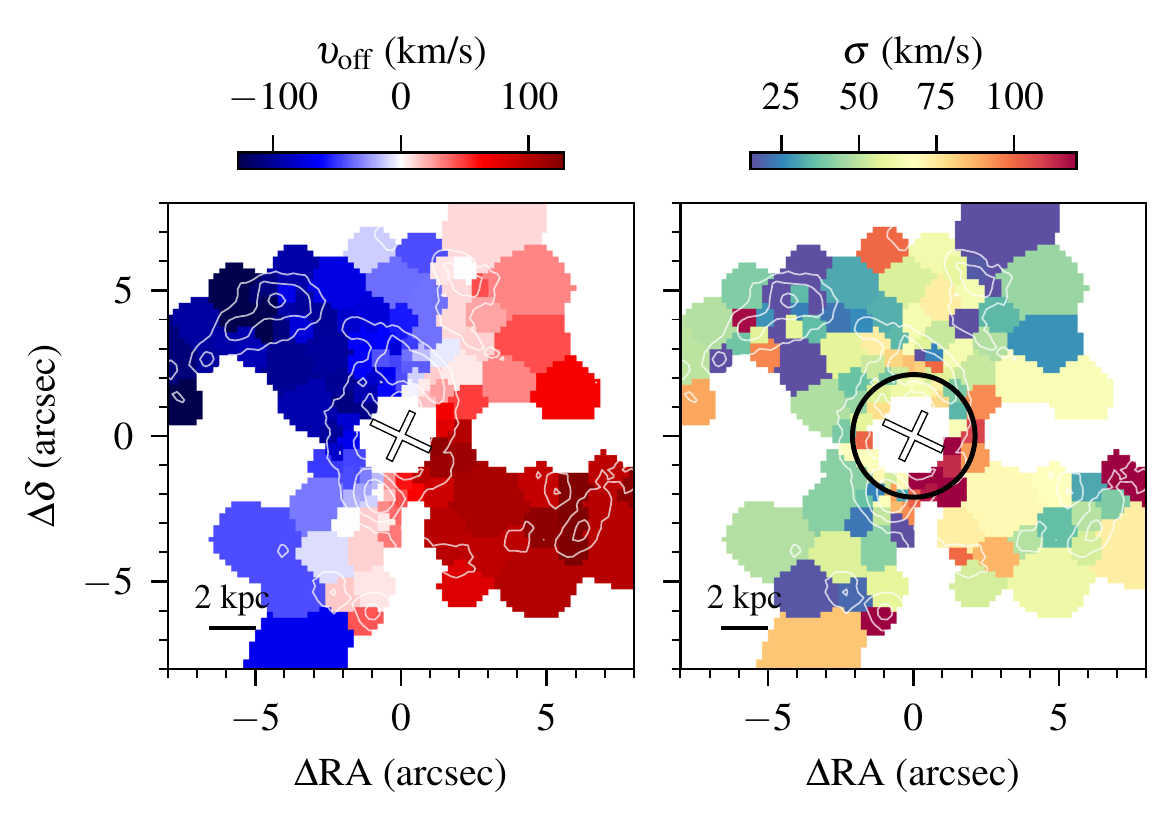}
  \includegraphics[width=\linewidth]{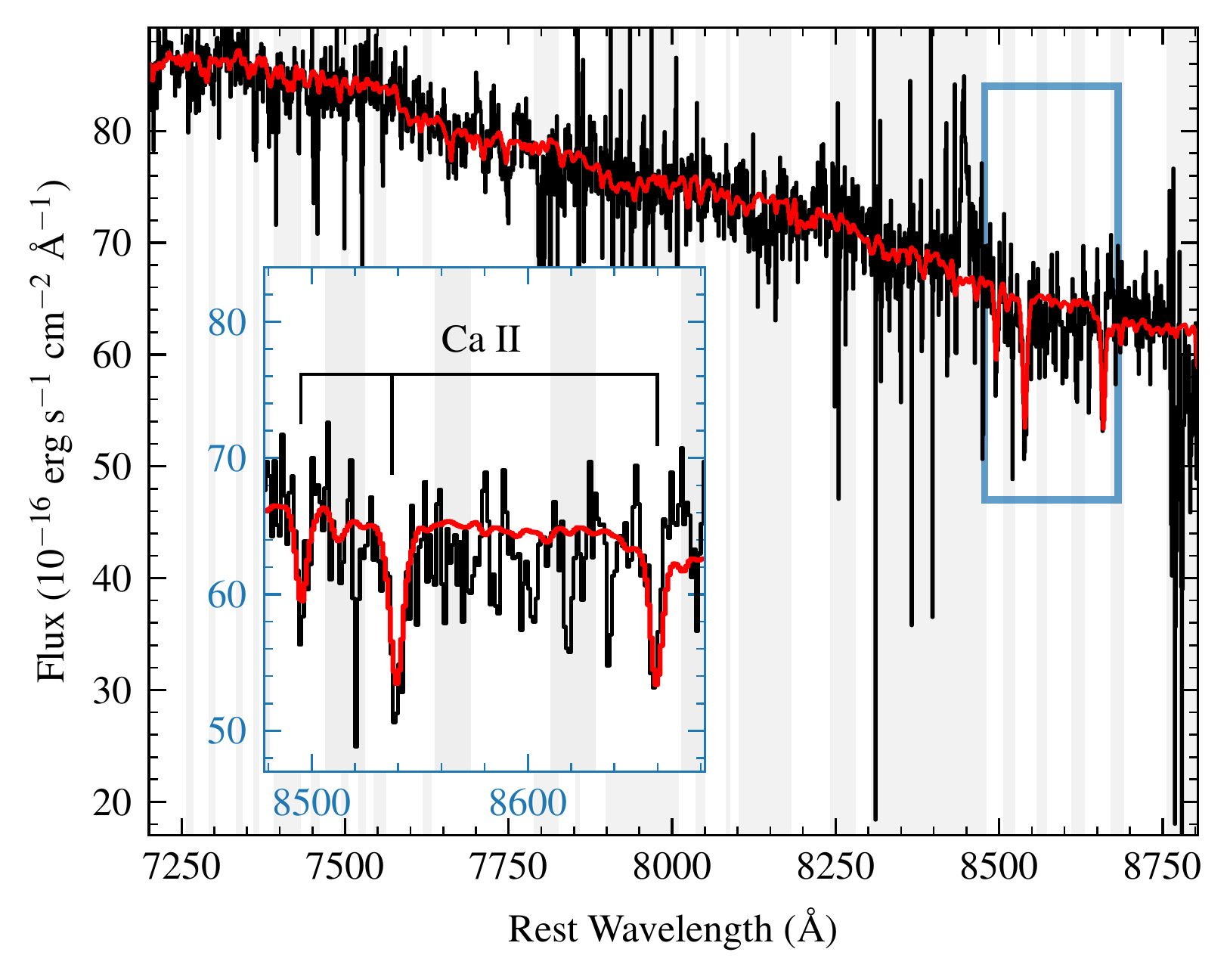}
  \caption{\textit{Top}: Stellar velocity (left) and velocity dispersion (right) maps of the host galaxy derived from the fit of the Ca\,\textsc{ii} triplet absorption features ($\lambda\lambda 8498, 8542, 8662$\,\AA). Only the bins with ${\chi}^2/{\rm d.o.f.} < 3$ are shown to exclude unreliable fits. The white cross marks the kinematic center and is oriented according to the position angle and inclination derived from the kinematic modeling shown in Fig.~\ref{fig:kinematic_model}. The black circle marks the projected aperture used to average the stellar velocity dispersion for the $M_{\rm BH}$--$\sigma_\star$ estimate. \textit{Bottom}: Example of the stellar kinematic fit for a Voronoi bin located about 1.5\,kpc northeast of the nucleus. Although the spectrum (solid black line) at red wavelengths is noisy and has forests of sky line subtraction residuals, the Ca\,\textsc{ii} triplet absorption features are robustly detected and fit with stellar population templates (red), allowing us to derive the stellar velocity and velocity dispersion in each bin. The inset shows a zoom-in on the \ion{Ca}{ii} triplet region. The best-fit model yields a stellar velocity of $v_\star \approx -90$ km s$^{-1}$ and a velocity dispersion of $\sigma_\star \approx 60$ km s$^{-1}$. The shaded gray regions indicate the masked ranges containing sky line subtraction residuals that were excluded from the fit.}
  \label{fig:stars_kinematics}
\end{figure}

The AGN-subtracted cube fitting procedure (Sect.~\ref{sec:emline_fitting}) yields spatially resolved maps of the velocity offset and velocity dispersion of the ionized interstellar medium across the galaxy (Fig.~\ref{fig:kinematic_model}, top panels). However, the PSF subtraction method also removes the continuum in the nuclear region, preventing a direct analysis of the stellar populations. To recover the stellar kinematics, we focused on the red part of the original cube, where the AGN PSF is smaller and the nuclear continuum less dominant.
We modeled the stellar continuum across the spectral range 7650--9350~\AA, corresponding to rest-frame stellar templates in the range $\sim$7200--8805 \AA\ given the redshift of the source.
This range includes the Ca\,\textsc{ii} triplet absorption lines ($\lambda8498$, $\lambda8542$, and $\lambda8662$\,\AA), which allow us to infer the stellar kinematics. The blue part of the spectrum is more affected by the AGN emission, making it less suitable for reliable stellar kinematic measurements in this case.
We adopted the E-MILES single-stellar population (SSP) model spectra (\citealt{Koleva:2012a,Vazdekis:2016a,LaBarbera:2017a}), which cover the spectral range 1680--50000 \AA.
The fit was performed using pPXF \citep{cappellari_full_2023}, applying the fit to Voronoi-binned spectra. 
To define the spatial bins, we computed the mean flux and its RMS within a red continuum window (8000--8250 \AA\ observed-frame) free from strong sky residuals, adopting a target S/N of 20 per spectral pixel.
This provides a good trade-off between the S/N required for reliable CaT absorption line measurements and spatial resolution. Sky line subtraction residuals, which persist as a constant background and accumulate when integrating multiple spaxels, were masked during the fit.
The resulting maps of stellar velocity and velocity dispersion are displayed in Fig.~\ref{fig:stars_kinematics}.
Reliable stellar velocity dispersions were obtained beyond $\sim$1.5\,kpc from the nucleus, while bins with ${\chi}^2/{\rm d.o.f.} > 3$ were discarded and masked in the above maps to limit contamination from spurious models. An example of the fit is shown in Fig.~\ref{fig:stars_kinematics}.

To investigate the gas dynamics and the possible presence of noncircular motions, we modeled the velocity fields with \texttt{DysmalPy} \citep{Lee2025}. We first fit the stellar velocity field to constrain the galaxy orientation. For the gas kinematics, we adopted a disk+bulge mass model (Fig.~\ref{fig:kinematic_model}), with the bulge effective radius constrained to $0.2-1.0$\,kpc, consistent with the bulge mass--size relation of \citet{Hashemizadeh2022} and the expected bulge mass from the local $M_{\rm BH}-M_{\rm bulge}$ relation. We then compared three kinematic scenarios: (i) pure circular rotation, (ii) rotation plus radially uniform inflow, and (iii) rotation with inflow along a bar potential. Inflow components did not significantly improve the fits relative to the pure rotation model or yield robust inflow velocities at the present spatial resolution. We discuss the implications for BH fueling in Sect.~\ref{sec:fueling}.

\section{Results}
\label{sec:results}
\subsection{Black hole mass constraints}

Estimating the black hole mass ($M_{\rm BH}$) in Ton S180 is crucial to assess its accretion regime. Lacking direct optical reverberation mapping, we use single-epoch (SE) virial mass estimators.
In these techniques, the black hole mass is computed using the virial theorem:
\begin{equation}
  M_{\rm BH} = f \frac{R_{\rm BLR} \Delta V^2}{G}
,\end{equation}
where $\Delta V$ is the velocity width of the broad line, measured as either the line dispersion ($\sigma$) or FWHM in the spectrum.
These calibrations use the empirical radius--luminosity ($R_{\rm BLR}$--$L$) relation, inferring the BLR radius from the ionizing luminosity \citep[e.g.,][]{Bentz2013}.
\citet{dalla_bonta_sloan_2020, dalla_bonta_estimating_2025} provide calibrations for both H$\alpha$ and H$\beta$ lines, including Eddington-ratio corrections, but they are calibrated mainly for moderate Eddington ratios ($-2 \lesssim \log\dot{m} \lesssim 0$), so their extension to super-Eddington rates remains uncertain.
\citet{woo_new_2026} recently presented H$\beta$ SE mass estimators for high-accretion AGNs, calibrated on super-Eddington sources with reverberation mapping. This relation accounts for the compact BLR expected in highly accreting AGNs \citep{du_radius_2019}, which can otherwise lead to underestimated BLR radii and black hole masses. We report results with both this calibration and the \cite{dalla_bonta_sloan_2020, dalla_bonta_estimating_2025} relations, which remain useful cross-checks. The latter infer the BLR radius from broad H$\beta$ and H$\alpha$ line luminosities to minimize host contamination, while \citet{woo_new_2026} uses the 5100\,\AA\ continuum luminosity, which is more directly related to the ionizing luminosity but can be more affected by host contamination. Given the extreme brightness of Ton S180 and the eigenspectrum decomposition discussed in Sect.~\ref{sec:agnspec}, we assume host contamination to be negligible in the nuclear spectrum.
For the virial factor $f$, we adopt $\langle \log f \rangle = 0.68 \pm 0.03$ for the relations in \citet{dalla_bonta_sloan_2020, dalla_bonta_estimating_2025}, $\log f = 0.05 \pm 0.12$ for the FWHM-based calibration and $\log f = 0.65 \pm 0.12$ for the $\sigma$-based calibration in \citet{woo_new_2026}, following the respective authors' prescriptions. This difference is expected, since $f$ depends on the adopted velocity-width definition and calibration scheme \citep{woo_new_2026}.
In Table~\ref{tab:bhmass}, we report the resulting mass estimates for Ton S180 based on the different calibrations and line-width definitions (see Table~\ref{tab:se_mass_breakdown} for a full parameter breakdown with and without the vBLR component).
As discussed by \citet{dalla_bonta_sloan_2020}, $\sigma$-based SE masses are generally more robust than FWHM-based ones, which can overestimate high masses and underestimate low masses. \citet{woo_new_2026} provide both $\sigma$- and FWHM-based relations and find no clear preference, but note that FWHM-based masses are less affected by strong wings. This is evident in Table~\ref{tab:bhmass}, where the two width definitions differ by 1.2 dex for their calibration.
The line luminosity to be used in the aforementioned relations was computed by integrating all fit components considered for the line, while the FWHM and $\sigma$ are derived numerically from the composite model profile using Model 3G (see Sect.~\ref{sec:agnspec}) as the best representation of the line profile.
The uncertainty on the derived mass is estimated by running the computation over the full set of 1000 Monte Carlo resampled best fits (see Sect.~\ref{sec:agnspec}), yielding a typical dispersion of $\sim$0.01 dex. To this we have to add in quadrature the intrinsic scatter of the SE relations (about 0.3 dex) and the uncertainty on the virial factor ($\sim$0.03 dex). The resulting total uncertainty is thus dominated by systematic errors, leading to a final uncertainty on the order of 0.3 dex on $\log M_{\rm BH}$.
In all of our mass estimates, we include the narrower Balmer component (nBLR) together with the broader BLR components when computing both line luminosity and line width, as its width ($\mathrm{FWHM} \approx 630\,\mathrm{km\,s^{-1}}$) is significantly broader than that of the host-galaxy narrow forbidden lines ($\mathrm{FWHM} \approx 300\text{--}380\,\mathrm{km\,s^{-1}}$).
Our high-S/N MUSE data also reveal the very broad Balmer component introduced in Sect.~\ref{sec:agnspec}, whose inclusion materially affects the SE mass range reported in Table~\ref{tab:bhmass}. The rationale for these choices, their comparison with previous work, and the impact of alternate profile decompositions are discussed in Appendix~\ref{appendix:agnmodel}.

\begin{table}[t]
\caption{Black hole mass estimates for Ton S180 using different methods.}
\tiny
\centering
\label{tab:bhmass}
\begin{tabular}{lcr}
\toprule
Method & $\log(M_{\rm BH}/M_\odot)$ & Reference \\
\midrule
SE H$\beta$ FWHM & $7.2 \pm 0.4$ & \cite{dalla_bonta_sloan_2020} \\
SE H$\beta$ $\sigma$ & $7.7 \pm 0.3$ & \cite{dalla_bonta_sloan_2020} \\
SE H$\alpha$ FWHM & $7.1 \pm 0.3$ & \cite{dalla_bonta_estimating_2025} \\
SE H$\alpha$ $\sigma$ & $7.6 \pm 0.2$ & \cite{dalla_bonta_estimating_2025} \\
SE H$\beta$ FWHM & $6.5 \pm 0.4$ & \cite{woo_new_2026} \\
SE H$\beta$ $\sigma$ & $7.7 \pm 0.4$ & \cite{woo_new_2026} \\

$M_{\rm BH}-\sigma_\star$ & $6.7 \pm 0.3$ &
\cite{kormendy_coevolution_2013} \\
\bottomrule
\end{tabular}
\tablefoot{
  The velocity measure used in the SE virial mass estimators was derived from the line second order moment ($\sigma$) or FWHM, as indicated using the spectral decomposition from Model 3G (see Sect.~\ref{sec:agnspec}) and including the contribution from the vBLR, iBLR, and nBLR components (see Table~\ref{tab:se_mass_breakdown} for a breakdown with and without the vBLR component). Including the vBLR yields $\sigma > \text{FWHM}$ because line dispersion is weighted quadratically by velocity offset, making it highly sensitive to broad wings. The last row reports the independent mass estimate from the $M_{\rm BH}-\sigma_\star$ relation using the stellar velocity dispersion measured from the \ion{Ca}{ii} triplet absorption lines in the AGN-subtracted cube (see Sect.~\ref{sec:kinematics_method}).}
\end{table}
\citet{scharwachter_spatially_2017} reported $\log(M_{\rm BH}/M_\odot) = 6.88 \pm 0.05$ from the H$\beta$ FWHM, without including the vBLR component (which was not detected in their data) and using the SE calibration from \cite{rakshit_catalog_2017}. This value is lower than our vBLR-inclusive estimates but higher than the black-hole mass obtained with the updated super-Eddington H$\beta$ FWHM calibration of \citet{woo_new_2026}.
Recent X-ray reverberation mapping of the Fe K$\alpha$ lag yields $\log(M_{\rm BH}/M_\odot) = 7.46_{-0.35}^{+0.01}$ \citep{kumar_reverberation_2026}, near the upper end of our range. This method models relativistic time delays between the primary coronal emission and the reflected disk component. Such X-ray-based estimates remain strongly model-dependent because they rely on assumptions about the coronal geometry and standard thin-disk structure, making them much less robust than optical reverberation mapping results.

An independent black hole mass estimate can be obtained from the stellar velocity dispersion ($\sigma_\star$) measured from the \ion{Ca}{ii} triplet absorption lines in the AGN-subtracted cube (see Sect.~\ref{sec:kinematics_method}). Although the nuclear region is contaminated by the AGN PSF, we can still measure $\sigma_\star$ in annuli around the nucleus where the stellar absorption features are detectable. Averaging the stellar velocity dispersion measured in the Voronoi bins within the projected 3\,kpc aperture marked in Fig.~\ref{fig:stars_kinematics} yields $\langle \sigma_\star \rangle = 80 \pm 3$\,km\,s$^{-1}$. Applying the $M_{\rm BH}$--$\sigma_\star$ relation from \citet{kormendy_coevolution_2013} gives a black hole mass estimate of $\log(M_{\rm BH}/M_\odot) = 6.7 \pm 0.3$, consistent with the SE virial estimates from the emission lines. This provides an independent, stellar-based constraint on the central black hole mass.
Our extensive tests highlight how black hole mass estimates in the absence of reliable optical reverberation mapping can be highly uncertain, especially for super-Eddington sources where the BLR structure and kinematics may deviate from the assumptions underlying standard SE calibrations.

\subsection{Accretion state and quasar main sequence placement}
\label{sec:Lbol}
To confirm the accretion regime of Ton S180, we estimated the dimensionless accretion rate $\dot{m}$ (see Sect.~\ref{sec:intro}, Eq.~\ref{eq:mdot}) directly from the optical luminosity and black hole mass, using the relation from \citet{du_fundamental_2016} calibrated for high-accretion rate AGNs:
\begin{equation}
\dot{m} = 20.1 \left(\dfrac{L_{5100}}{10^{44} \text{erg\,s}^{-1} \cos i}\right)^{3/2} \left(\dfrac{M_{\rm BH}}{10^7\,M_\odot}\right) ^{-2} \, ,
\end{equation}
where $L_{5100} \equiv \lambda L_\lambda(5100\,\text{\AA})$ is the monochromatic luminosity at 5100\,\AA\ in erg\,s$^{-1}$ and $\cos i$ accounts for the disk inclination. For Ton S180, we adopt $\cos i = 0.75$, corresponding to $i \approx 41^\circ$, as commonly assumed for type 1 AGNs \citep{du_fundamental_2016}.
Using our measured $L_{5100} = 2.03 \times 10^{44}$\,erg\,s$^{-1}$, corrected only for Galactic extinction, we derive a wide range of accretion rates depending on the allowed black-hole mass. Across the interval retained in Table~\ref{tab:bhmass}, from $\log(M_{\rm BH}/M_\odot)=6.5$ to $7.7$, the dimensionless accretion rate spans from $\dot{m} = 1000 \pm 200$ to $\dot{m} = 4 \pm 1$.
The lower end of the mass range therefore implies an extreme slim-disk regime, whereas the upper end still indicates a highly accreting source. We then estimated the expected Eddington ratio, $\lambda_{\rm Edd}$, from $\dot{m}$ using the relation from \citet{madau_super_2014}
\begin{equation}
  \lambda_{\rm Edd} = A(a) \left[ \frac{0.985}{16/\dot{m} + B(a)} + \frac{0.015}{16/\dot{m} + C(a)} \right],
\end{equation}
where $A(a)$, $B(a)$, and $C(a)$ are the spin-dependent coefficients given by \citet{madau_super_2014}; we adopted zero spin as a fiducial choice. For the minimum mass, we obtain $\lambda_{\rm Edd}^{\rm sup} = 4.08^{+0.01}_{-0.06}$, and for the maximum mass $\lambda_{\rm Edd}^{\rm sup} = 2.9 \pm 0.7$. In both cases, Ton S180 remains super-Eddington. Inverting Eq.~\ref{eq:lamba_Edd} then gives the bolometric luminosities reported in Table~\ref{tab:eddington}, which reflect the order-of-magnitude uncertainty on the black hole mass.
For comparison, we also computed the bolometric luminosity and Eddington ratio using the standard approach for lower accretion rate AGNs. We adopted the log-linear bolometric correction from \citet{runnoe_updating_2012} for $5100\,\text{\AA}$:
\begin{equation}
    \log (L_{\rm iso}/{\rm erg\,s^{-1}}) = 4.89 + 0.91 \log (L_{5100}/{\rm erg\,s^{-1}}) \, .
\end{equation}
This correction is calibrated for moderate Eddington ratios and is likely not appropriate for super-Eddington AGNs, but it provides a useful comparison. Applying the anisotropy correction $L_{\rm bol} \approx 0.75 \times L_{\rm iso}$ \citep{runnoe_updating_2012}, we obtain
\begin{equation*}
  L_{\rm bol}^{5100} = (1.2 \pm 0.3) \times 10^{45}\,\mathrm{erg\,s}^{-1}. 
\end{equation*}
The resulting thin-disk Eddington ratios, $\lambda_{\rm Edd}^{5100} = L_{\rm bol}^{5100}/L_{\rm Edd}$, span $3 \pm 3$ at the low-mass end and $0.21 \pm 0.16$ at the high-mass end. Under this prescription, the source could be either super-Eddington or mildly sub-Eddington depending on the adopted black hole mass. Despite the differences between prescriptions, Ton S180 is unambiguously in a high-accretion state; the uncertainty concerns how extreme the accretion is.

\begin{table*}[ht]
  \caption{Accretion properties of Ton S180 for the minimum and maximum black-hole masses retained in Table~\ref{tab:bhmass}.}
\tiny
\centering
\label{tab:eddington}
\begin{tabular}{l | ccc | cc}
\toprule
$\log (M_{\rm BH}/M_\odot)$ & $\dot{m}$ & $\lambda_{\rm Edd}^{\rm sup}$ & $L_{\rm bol}^{\rm sup}$ (erg s$^{-1}$) & $\lambda_{\rm Edd}^{5100}$ & $L_{\rm bol}^{5100}$ (erg s$^{-1}$) \\
\midrule
$6.5 \pm 0.4$ & $1000 \pm 200$ & $4.08^{+0.01}_{-0.06}$ & $(1.6 \pm 0.1) \times 10^{45}$ & $3 \pm 3$ & $(1.2 \pm 0.3) \times 10^{45}$ \\
$7.7 \pm 0.3$ & $4 \pm 1$ & $2.9 \pm 0.7$ & $(1.8 \pm 0.4) \times 10^{46}$ & $0.21 \pm 0.16$ & $(1.2 \pm 0.3) \times 10^{45}$\\
\bottomrule
\end{tabular}
\tablefoot{$\dot{m}$ was computed from the spectrum following the \citet{du_fundamental_2016} prescription. The corresponding Eddington ratios were computed with the \citet{madau_super_2014} theoretical relation. The fourth column reports the bolometric luminosity obtained inverting the Eddington ratio definition.
The last two columns report the Eddington ratio and the bolometric luminosity obtained using the \citet{runnoe_updating_2012} relation, which is, however, calibrated for moderate Eddington ratios.}
\end{table*}

\begin{figure}
  \centering
  \includegraphics{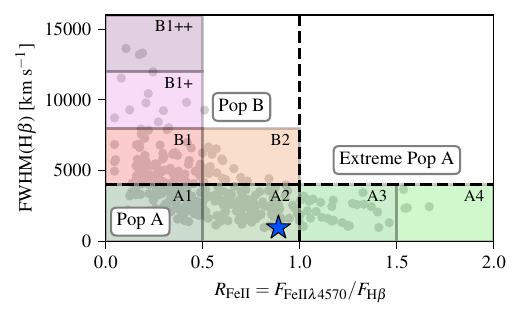}
  \caption{Quasar MS diagram. Ton S180 is plotted as a blue star. The source falls within population A. The data points plotted in the background are taken from \cite{marziani_sulentic_highly_2014}.}
  \label{fig:quasarMS}
\end{figure}

The optical iron strength provides an independent empirical view of the same accretion regime. Quasar spectral diversity is organized along the ``main sequence'' (MS) of Eigenvector 1 \citep{boroson_emission-line_1992}, defined by FWHM(H$\beta$) and the optical iron strength $R_{\rm \ion{Fe}{ii}} = F_{\rm \ion{Fe}{ii}\lambda4570}/F_{\rm H\beta}$, where \ion{Fe}{ii}$\lambda4570$ denotes the integrated iron emission over 4434--4684 \AA. Driven primarily by the Eddington ratio, it separates lower-accreting Population B from high-accreting Population A sources (FWHM(H$\beta$) $\le 4000$ km\,s$^{-1}$; \citealt{boroson_emission-line_1992, marziani_superedd_2025}), with super-Eddington systems expected to cluster in the ``Extreme Population A'' domain ($R_{\rm \ion{Fe}{ii}} \ge 1$). High Eddington ratios favor a denser and less ionized BLR, which strengthens \ion{Fe}{ii} relative to H$\beta$ \citep{marziani_superedd_2025}.
For Ton S180, integrating the \ion{Fe}{ii} blend over 4434--4684 \AA\ and normalizing by the total broad H$\beta$ flux of our decomposition, we derive $R_{\rm \ion{Fe}{ii}} \simeq 0.89$. This places the source in the spectral bin A2 of the MS, as shown in Fig.~\ref{fig:quasarMS}. Although it does not cross the $R_{\rm \ion{Fe}{ii}} \ge 1$ threshold typical of extreme accretors, its location is consistent with the regime where $\lambda_{\rm Edd}$ steadily increases before saturating around $R_{\rm \ion{Fe}{ii}} \approx 0.84$ \citep{marziani_superedd_2025}. Our value is consistent with the $R_{\rm \ion{Fe}{ii}} \sim 0.9$ reported by \citet{scharwachter_spatially_2017}.

We also observe a kinematic gradient across the BLR, where broader components exhibit progressively larger systemic blueshifts, from the mildly outflowing nBLR (about 50 km s$^{-1}$) to the faster vBLR ($>500$ km s$^{-1}$). This velocity stratification is consistent with an outflowing BLR driven by radiation pressure, in which the narrower components trace the denser, slower base and the broadest component traces the accelerating, optically thinner phases \citep[e.g.,][]{murray_accretion_1995}. It further supports the high-Eddington regime in Ton S180.

\subsection{Host-galaxy morphology and kinematics}
\label{sec:host}
The AGN-subtracted H$\alpha$ and [\ion{N}{ii}] maps reveal a prominent circumnuclear ring connecting the spiral arms together with an inner elongated NE--SW structure compatible with a bar, with semi-major axis $\gtrsim 1$ kpc (Fig.~\ref{fig:fluxmaps}). Fitting the stellar velocity field yields a position angle of PA $= 64^\circ \pm 2^\circ$ and a line-of-sight inclination of $i = 37^\circ \pm 2^\circ$.
The pure circular rotation model already provides a good description of the large-scale gas velocity field, but it does not fully reproduce the data. Residuals reach up to $\sim 50$\,km\,s$^{-1}$ in both velocity and velocity dispersion. Including inflow components yields negligible inflow velocities ($<10$\,km\,s$^{-1}$) and does not significantly improve the residuals. Beyond $\sim 2$\,kpc, the kinematics remain rotation-dominated, yet coherent residuals persist along the spiral-arm region seen in Figs.~\ref{fig:CubeImage} and \ref{fig:fluxmaps}, consistent with non-axisymmetric streaming motions expected in barred spirals \citep{canzian_corotation_1993, sempere_pattern_1995}.
Inside $\sim 2$\,kpc, within the circumnuclear ring, the residual pattern becomes less ordered and the velocity-dispersion residuals remain strong (Fig.~\ref{fig:kinematic_model}). Restricting the inflow fit to this inner region still does not return significant inflow velocities or lower residuals.

\subsection{Ionized outflow energetics}
Estimating the physical properties of the ionized outflow requires one to derive the electron density ($n_e$) and mass ($M_{\rm ion}$) of the emitting gas. To derive the energetics, we adopted the fluid relation framework described in \citet{carniani_ionised_2015} and \citet{cresci_blowin_2015}.
We distinguished between the unresolved nuclear outflow and the extended outflow detected in the AGN-subtracted cube. For the unresolved component, we used the broad [\ion{O}{iii}] wing seen in the PSF-integrated spectrum as our conservative estimate and also considered the full [\ion{O}{iii}] profile as a less conservative alternative. Comparing these components allowed us to assess how efficiently the AGN couples mechanical power to the surrounding gas through the acceleration of galactic-scale outflows.

\subsubsection{The nuclear outflow}
For the unresolved nuclear wind, we measure an outflow luminosity of $L_{[\ion{O}{iii}]} = (1.6 \pm 0.2) \times 10^{41}$\,erg\,s$^{-1}$ and a maximum velocity of $v_{\rm max} = |v_\mathrm{wing}| + 2\sigma_\mathrm{wing} = 1150 \pm 50$\,km\,s$^{-1}$ \citep[e.g.,][]{Rupke2013, carniani_ionised_2015}, where $v_\mathrm{wing}$ and $\sigma_\mathrm{wing}$ are the offset velocity and the velocity dispersion of the most blueshifted Gaussian component of [\ion{O}{iii}] in our nuclear fitting (see Table~\ref{tab:AGNfit}). As direct density diagnostics are not available for the unresolved outflow, we assumed a typical outflow electron density of $n_e = 500$\,cm$^{-3}$, consistent with values commonly adopted for ionized AGN winds \citep{davies_ionized_2020, carniani_ionised_2015,Venturi2026}.
We did not apply an internal dust extinction correction to the nuclear [\ion{O}{iii}] luminosity because deblending the narrow Balmer lines from the intense broad-line profile (iBLR, vBLR) and \ion{Fe}{ii} multiplets introduces strong systematic degeneracies in the nuclear Balmer decrement, making our nuclear mass outflow rate and kinetic power conservative lower bounds. Given that Ton S180 is a bare NLSy1 with negligible intrinsic dust extinction \citep{turner_chandra_2001}, any nuclear reddening is expected to be modest.
Assuming solar metallicity and using the calibration from \citet{carniani_ionised_2015}, we estimated the total ionized gas mass:
\begin{equation}
\label{eq:Mion}
  M_{\rm ion} = 4.0 \times 10^7 \left( \frac{L_{[\ion{O}{iii}]}}{10^{44}\,{\rm erg\,s}^{-1}} \right) \left( \frac{n_e}{10^3\,{\rm cm}^{-3}} \right)^{-1} \, M_\odot.
\end{equation}
This yields $M_{\rm ion} = (1.27 \pm 0.18) \times 10^5\,M_\odot$.

Since this component is spatially unresolved, we adopted a maximum outflow radius equal to the half-width at half maximum (HWHM = FWHM/2) of the MUSE PSF, $R_{\rm out} \approx 0.3$\,kpc. Under the assumption of a conically averaged flow, the mass outflow rate is given by \citep[see e.g.,][]{cresci_blowin_2015}
\begin{equation}
\label{eq:Mdot}
    \dot{M}_{\rm out} = 3 M_{\rm ion} v_{\rm max} / R_{\rm out} \, .
\end{equation}
This yields a nuclear mass outflow rate of $\dot{M}_{\rm nucl} = 1.5 \pm 0.2\,M_\odot\,\mathrm{yr}^{-1}$. The associated kinetic power is $\dot{E}_{\rm kin} = (6.1 \pm 1.5) \times 10^{41}$\,erg\,s$^{-1}$. Comparing this to the bolometric luminosities derived in Sect.~\ref{sec:Lbol}, the kinetic coupling efficiency is on the order of $0.05$\%.

As a less conservative alternative, we also considered a scenario in which the entire nuclear [\ion{O}{iii}] profile is involved in the nuclear outflow. This profile combines the more strongly blueshifted wing component with the narrow core, which is itself blueshifted by $-170$~km\,s$^{-1}$ relative to the common systemic velocity and to all the other narrow lines in the spectrum. For this non-Gaussian, multicomponent profile, we adopted a non-parametric approach and defined the maximum outflow velocity as the second-percentile velocity offset ($v_{02} = 1070 \pm 50$\,km\,s$^{-1}$; e.g., \citealt{Liu2013}). Summing the line fluxes of the two components yields a total nuclear [\ion{O}{iii}] outflow luminosity of $L_{[\ion{O}{iii}], \text{tot}} = (3.6 \pm 0.3) \times 10^{41}$\,erg\,s$^{-1}$ and an ionized mass of $M_{\rm ion, tot} = (2.9 \pm 0.2) \times 10^5\,M_\odot$. Under the same conical flow framework from Eq.~\ref{eq:Mdot}, this gives a higher nuclear mass outflow rate of $\dot{M}_{\rm nucl, tot} = 3.2 \pm 0.4\,M_\odot\,\mathrm{yr}^{-1}$ and a kinetic power of $\dot{E}_{\rm kin, tot} = (1.2 \pm 0.3) \times 10^{42}$\,erg\,s$^{-1}$ ($\dot{E}_{\rm kin}/L_{\rm bol} \approx 0.095$\%). The physical parameters derived for all nuclear and extended outflow configurations are summarized in Table~\ref{tab:outflow_summary}.

\begin{table*}
\caption{Derived physical and kinematic parameters for the nuclear and extended ionized outflows in Ton~S180.}
\label{tab:outflow_summary}
\centering
\tiny
\begin{tabular}{@{\hspace{2pt}}l@{\hspace{4pt}}c@{\hspace{4pt}}c@{\hspace{4pt}}c@{\hspace{8pt}}c@{\hspace{8pt}}c@{\hspace{8pt}}c@{\hspace{2pt}}}
\toprule
Outflow component & $L_{[\ion{O}{iii}]}$ & $M_{\rm ion}$ & $v_{\rm max}$ & $R_{\rm out}$ & $\dot{M}_{\rm out}$ & $\dot{E}_{\rm kin}$ \\
 & ($10^{40}$\,erg\,s$^{-1}$) & ($M_\odot$) & (km\,s$^{-1}$) & (kpc) & ($M_\odot$\,yr$^{-1}$) & (erg\,s$^{-1}$) \\
\midrule
Unresolved nuclear (wing only) & $16 \pm 2$ & $(1.27 \pm 0.18) \times 10^5$ & $1150 \pm 50$ & $<0.3$ & $>1.5 \pm 0.2$ & $>(6.1 \pm 1.5) \times 10^{41}$ \\
Unresolved nuclear (total [\ion{O}{iii}]) & $36 \pm 3$ & $(2.9 \pm 0.2) \times 10^5$ & $1070 \pm 50$ & $<0.3$ & $>3.2 \pm 0.4$ & $>(1.2 \pm 0.3) \times 10^{42}$ \\
Extended & $0.101$ & $8.1 \times 10^2$ & $340 \pm 10$ & 2.0 & $4.2 \times 10^{-4}$ & $1.5 \times 10^{37}$ \\
\bottomrule
\end{tabular}
\tablefoot{
All outflow calculations assume an electron density $n_e = 500$\,cm$^{-3}$ and the conically averaged flow prescription $\dot{M}_{\rm out} = 3 M_{\rm ion} v_{\rm max} / R_{\rm out}$ (Eq.~\ref{eq:Mdot}). For the unresolved nuclear outflow, $R_{\rm out} \approx 0.3$\,kpc corresponds to the MUSE PSF HWHM.
}
\end{table*}

\subsubsection{The extended outflow}
The spatially resolved outflow extends up to $R_{\rm out} \sim 2$\,kpc west of the nucleus (Fig.~\ref{fig:outflow}). It is significantly fainter than the unresolved nuclear component, with an intrinsic [\ion{O}{iii}] luminosity of $L_{[\ion{O}{iii}]} \approx 1.0 \times 10^{39}$\,erg\,s$^{-1}$, corrected for internal dust extinction. The color excess in the outflow aperture was estimated from the narrow-line Balmer decrement \citep{osterbrock_astrophysics_2006} assuming intrinsic Case B recombination ($\mathrm{H}\alpha/\mathrm{H}\beta = 2.85$), yielding $E(B-V) \approx 0.78$\,mag ($A_{[\ion{O}{iii}]} \approx 2.7$\,mag).
The outflow kinematics are also more moderate than in the nucleus, with a maximum velocity of $v_{\rm max} = 340 \pm 10$\,km\,s$^{-1}$ ($380 \pm 10$\,km\,s$^{-1}$ relative to the local host-galaxy disk centroid). At the outflow radius, the systemic gas has projected kinematics $|v| \sim 140$\,km\,s$^{-1}$ and $\sigma \sim 50$\,km\,s$^{-1}$ from the host modeling (Sect.~\ref{sec:kinematics_method}). In a truncated halo, $v_{\rm esc}(r) \simeq [2(1+\ln(r_{\rm max}/r))]^{1/2} v_{\rm c}$ \citep{binney_galbook_2008}; taking the observed local rotation amplitude as an order-of-magnitude proxy for $v_{\rm c}$ and $r_{\rm max} \sim 10$\,kpc gives $v_{\rm esc} \sim 300$\,km\,s$^{-1}$ at $r \simeq 1.3$\,kpc. The gas can therefore plausibly escape the inner bulge.
Since outflow signatures appear only in [\ion{O}{iii}], other line diagnostics (such as [\ion{S}{ii}]) trace the galactic disk rather than the wind. We therefore adopt the same $n_e = 500$\,cm$^{-3}$ used for the unresolved nuclear outflow and apply the \citet{carniani_ionised_2015} relation (Eq.~\ref{eq:Mion}) consistently for both nuclear and extended components.
This yields an ionized gas mass of $M_{\rm ion} \approx 8.1 \times 10^{2}\,M_\odot$. Using Eq.~\ref{eq:Mdot} under the same conical flow assumption, the resulting mass outflow rate is $\dot{M}_{\rm ext} \approx 4.2 \times 10^{-4}\,M_\odot\,\mathrm{yr}^{-1}$ with a kinetic power of $\dot{E}_{\rm kin} \approx 1.5 \times 10^{37}$\,erg\,s$^{-1}$. For comparison, using the uncorrected [\ion{O}{iii}] flux ($L_{[\ion{O}{iii}]} \approx 8.5 \times 10^{37}$\,erg\,s$^{-1}$) provides lower bounds of $M_{\rm ion} \approx 6.8 \times 10^1\,M_\odot$, $\dot{M}_{\rm ext} \approx 3.5 \times 10^{-5}\,M_\odot\,\mathrm{yr}^{-1}$, and $\dot{E}_{\rm kin} \approx 1.3 \times 10^{36}$\,erg\,s$^{-1}$.

\subsubsection{Comparison with the X-ray UFO}
To place the ionized outflow in a broader radial feedback picture, we compare the X-ray UFO with both the unresolved nuclear [\ion{O}{iii}] outflow and the faint extended [\ion{O}{iii}] component. While the momentum rate $\dot{P}=\dot{M}v$ traces the thrust imparted to the ambient gas, $\dot{E}_{\rm kin}=\dot{P}v/2$  traces how much energy is carried by the outflow in its propagation. \citet{matzeu_first_2020} report a detection of blueshifted absorption from \ion{O}{viii} and \ion{C}{vi} associated with a nuclear UFO in Ton S180 with $v_{\rm out}\approx0.2c$, $\log\xi\sim2.7$, and $N_{\rm H}<10^{22}$\,cm$^{-2}$. With these values we can estimate the UFO properties using the prescription of \citet{tombesi_unification_2013},
\begin{equation}
\dot{M}_{\rm out} = 4\pi C_f\mu m_p r_{\rm wind} N_H v_{\rm out},
\end{equation}
where $C_f=0.3$ is an average covering fraction of the wind as suggested by population studies \citep[e.g.,][]{tombesi_evidence_2010}, $\mu=1.4$ is the mean molecular weight, and $r_{\rm wind}=r_{\rm min}=2GM_{\rm BH}/v_{\rm out}^2$ is the launching radius of the wind, which we conservatively assume to coincide with the escape radius at the observed velocity. With these assumptions, we obtain
\begin{align*}
  & \dot{M}_{\rm UFO}\lesssim(0.18 \div 2.9) \times10^{-3} \,M_\odot\,\mathrm{yr}^{-1} \\
  & \dot{E}_{\rm UFO}\lesssim(5 \div 260) \times10^{40} \, \text{erg s}^{-1},
\end{align*}
for $\log(M_{\rm BH}/M_\odot)=6.5 \div 7.7$. These values correspond to $\dot{E}_{\rm UFO}/L_{\rm bol}\lesssim0.004 \div 0.25\%$. The corresponding UFO momentum rate is $\dot{P}_{\rm UFO} \sim 10^{33}$ dyn, to be compared with $\dot{P}_{\rm nucl} \sim (1.0 \text{--} 2.0) \times 10^{34}$ dyn for the unresolved nuclear [\ion{O}{iii}] outflow (spanning the conservative wing-only and total [\ion{O}{iii}] profile estimates; Table~\ref{tab:outflow_summary}) and $\dot{P}_{\rm ext}\approx 10^{30}$ dyn for the resolved component. All three estimates remain below $L_{\rm bol}/c \sim 3 \times 10^{34}$ dyn, but the momentum flux drops by almost four orders of magnitude from the unresolved to the resolved ionized phase. Because $N_H$ is an upper limit and $r_{\rm min}$ is a lower-limit radius, these estimates remain indicative. Even so, the UFO kinetic power overlaps the estimate for the unresolved nuclear [\ion{O}{iii}] outflow, while both phases exceed the extended ionized component by several orders of magnitude. The emerging picture is therefore strongly scale dependent, and it does not favor a simple one-to-one transfer of mechanical power to the extended ionized outflow. At the same time, the comparison with the resolved component should be treated with caution, because the kiloparsec-scale gas may still be responding to an earlier accretion state rather than to the current high-accretion episode.

\section{Discussion}
\label{sec:discussion}

\subsection{Ton S180 in the context of highly accreting NLSy1s}
The nuclear properties of Ton S180 place it among the most rapidly accreting local NLSy1s, although the inferred accretion rate and derived Eddington ratio remain sensitive to the adopted black hole mass calibration. The inferred SE black hole mass indicates high accretion rates, with $\dot{m} \sim 4\text{--}1000$, while the optical iron strength of $R_{\rm \ion{Fe}{ii}} \simeq 0.89$ places the source in the Population A domain of the quasar MS.
Ton S180 also illustrates the difficulty of measuring black hole masses in extreme-accretion Seyferts. Optical reverberation mapping is available for only a modest subset of high-accretion AGNs compared with the much larger NLSy1 population, and the SEAMBH campaigns have shown that high-$\dot{m}$ systems often display H$\beta$ lags shorter than expected from the canonical broad-line region size--luminosity relation \citep{du_fundamental_2016, du_radius_2019}. This behavior is commonly interpreted in terms of slim-disk radiative transfer effects, including anisotropy, self-shadowing, and photon trapping, which reduce the ionizing flux seen by part of the broad-line region. Ton S180 lacks an optical reverberation measurement, so its SE mass estimates remain structurally uncertain. The recent \citet{woo_new_2026} calibrations are especially relevant because they explicitly account for the dependence on Eddington ratio by using reverberation-mapped high-accretion AGNs, but even then the ambiguity remains severe: in Ton S180, the FWHM-based estimator gives $\log(M_{\rm BH}/M_\odot) \simeq 6.5$, whereas the velocity-dispersion-based estimator gives $\log(M_{\rm BH}/M_\odot) \simeq 7.7$. This $\sim 1.2$\,dex discrepancy is largely driven by the distinct line profile of Ton~S180. While the \citet{woo_new_2026} calibrations are centered on a reverberation-mapped reference sample with a mean ratio of $\mathrm{FWHM}/\sigma_{\rm mean} \approx 2.02 \pm 0.57$, Ton~S180 displays a significantly lower ratio of $\mathrm{FWHM}/\sigma \approx 0.66$ when the very broad component is included (Table~\ref{tab:se_mass_breakdown}). Moving away from the calibration mean profile shape systematically diverges the $\mathrm{FWHM}$- and $\sigma$-based mass estimates, underscoring how profile stratification, orientation, or non-virial broad wings can bias SE relations. Ton S180 therefore reinforces the broader conclusion that SE masses in the highest-accretion NLSy1s remain intrinsically uncertain without direct reverberation mapping or direct dynamical estimates through BLR interferometry. Indeed, SE mass estimates are likely consistently overestimated in highly accreting AGNs, as indicated by RM studies (due to the smaller broad-line region radius compared to the canonical size--luminosity relation) and recent interferometric studies \citep{GRAVITY_2024}.

On the outflow phenomenology side the picture is still heterogeneous. Unresolved optical spectroscopy has long shown that NLSy1s frequently host completely blueshifted and strongly asymmetric [\ion{O}{iii}] profiles, unlike other lines, usually interpreted as nuclear ionized outflows linked to high-Eddington accretion \citep[e.g., I\,Zw\,1 and the PUMA sample][]{Perna2021}. However, unresolved spectroscopy alone cannot establish whether the nuclear wind couples efficiently to the host galaxy ISM to drive extended galactic-scale outflows.
High-resolution spatially resolved optical studies of highly accreting NLSy1s remain largely unexplored, and Mrk~1044 is one of the very few clear comparison cases currently available. The spectroastrometric analysis of the MUSE narrow field mode data of this highly accreting NLSy1 ($\lambda_{\rm Edd}=0.96$) resolved a multiphase outflow over only a few parsecs, which is weak in terms of both velocity ($v \sim 211$ km s$^{-1}$) and mass outflow rate ($\dot{M}_{\rm out} \sim 0.002 M_\odot$ yr$^{-1}$), embedded in circumnuclear star formation \citep{winkel_close_2023}. A more powerful outflow is observed in the NLS1 IRAS~17020+4544 \citep{bellocchi_multiphase_2026}, which accretes at a slightly lower rate ($\lambda_{\rm Edd} \sim 0.7$) but is classified as a luminous infrared galaxy (LIRG) and hosts a low-power radio jet that may contribute to driving the wind. In this object, optical IFU data reveal fast ($v_{\rm out} \sim 1240-1460$ km s$^{-1}$) and spatially extended ($R_{\rm out} \sim 0.5-1$ kpc) warm ionized outflows with significant mass outflow rates ($\dot{M}_{\rm out} \sim 10-50 M_\odot$ yr$^{-1}$), nested inside a larger molecular outflow \citep{bellocchi_multiphase_2026}. Furthermore, the luminous quasar PDS~456 provides a higher-luminosity benchmark. Although formally classified as a Population A quasar rather than a NLS1, it shares similar accretion physics and is heavily super-Eddington ($\lambda_{\rm Edd} = 4.64$, dimensionless accretion rate $\dot{m} \sim 200$, according to recent direct dynamical mass measurements by \citealt{GRAVITY_2024}). However, PDS~456 differs significantly from our target: it is the most luminous radio-quiet AGN in the local Universe ($L_{\rm bol} \sim 10^{47}$ erg s$^{-1}$), resides in a galaxy overdensity \citep{Bischetti2019}, and drives a $\sim 12$ kpc [\ion{O}{iii}] outflow \citep{travascio_pds_2024}.
In this context, Ton S180 appears to be a case in which the transition between a strong nuclear wind and a weak host-scale ionized response can be resolved.

From a multiwavelength perspective, NLSy1s are among the AGN classes with the highest incidence of X-ray UFOs, and their UFO properties appear to bridge those of lower-luminosity Seyferts and luminous quasars in terms of outflow velocities, velocity dispersions, and column densities \citep{laurenti_ufo_2026}. Multi-epoch X-ray studies of individual archetypes, such as IRAS\,13224--3809, further underscore the highly dynamic nature of these central engines, which exhibit steeper X-ray spectra, more rapid flux variability on hour-long timescales, and a higher frequency of complex UFOs compared to standard broad-line AGNs \citep{Condo2026, laurenti_ufo_2026}. Ton\,S180 thus fits a broader scenario in which rapidly accreting AGNs can launch powerful compact winds \citep{matzeu_first_2020}. However, the efficiency with which those winds propagate into the resolved ionized medium remains largely unknown due to the limited number of studies. This coupling efficiency likely depends not only on accretion rate, but also on luminosity, duty cycle, gas phase, host structure, and environment. Building robust statistics of spatially resolved studies of highly accreting systems such as Ton\,S180 is therefore essential to understand these still uncertain and debated aspects of AGN feedback.

\subsection{Fueling BH accretion from the host galaxy}
\label{sec:fueling}
The circumnuclear ring may trace secular redistribution of gas angular momentum driven by non-axisymmetric perturbations. In barred galaxies, gas can accumulate near gravitational resonances \citep{Athanassoula1992, Sormani2023} and subsequently migrate inward \citep{Shlosman1989, Combes2023}, potentially helping sustain the super-Eddington BH accretion, but rings alone do not demonstrate current feeding \citep{buta_rings_1996, buta_rings_revisited_2017}. In Ton S180, the bar-like inner elongation and the residuals along the spiral-arm region are consistent with secular transport, but we cannot conclude whether inflowing motions are present within the circumnuclear ring, toward the nucleus. The failure of both uniform and bar-driven inflow kinematic models indicates that the inner gas kinematics is more complex than these simple prescriptions and likely combines inflow, outflow, and other non-axisymmetric motions. At the current spatial resolution of $\sim 1$\,kpc, a more definitive interpretation remains out of reach, and higher-resolution data are required to establish whether and how gas is transported from the ring to the nucleus.

\subsection{Feedback from the nucleus to kiloparsec scales}
The outflow picture is markedly scale dependent. In the unresolved nucleus, the [\ion{O}{iii}] wing traces a clear ionized outflow with sub-kiloparsec extent and measurable kinetic power, whereas the resolved component is orders of magnitude weaker in luminosity, mass outflow rate, and kinetic power. The absolute coupling depends strongly on the adopted density, which is not directly constrained by our data and may vary by orders of magnitude \citep[e.g.,][]{venturi_complex_2023}. Varying $n_e$ within 10--1000\,cm$^{-3}$ (Appendix~\ref{appendix:density}) broadens the kinetic coupling efficiency, $\dot{E}_{\rm kin}/L_{\rm bol}$, from $\sim0.03$\% to $\sim3$\% for the nuclear outflow and from $\sim10^{-6}$\% to $\sim10^{-4}$\% for the extended component.
Even so, the decline with radius remains evident.
The often adopted benchmark for efficient AGN feedback needed to quench star formation in the host is 0.5--5\% \citep{king_powerful_2015}, although nearby Seyfert studies suggest that feedback-relevant values can extend down to $\sim0.1$--0.5\% in some systems \citep{crenshaw_feedback_2012}. Furthermore, cosmological simulations often assume that a large fraction of the injected AGN energy is directly converted into outflow kinetic energy, whereas hydrodynamic models show that thermal losses and work against the potential reduce the kinetic energy transferred to the large-scale wind to $\lesssim 10\text{--}20\%$ \citep[e.g.,][]{richings_cooling_2018, Harrison2018}, effectively lowering the required kinetic coupling efficiency threshold.
This may mean that the wind dissipates within the inner few kiloparsecs or that the optical phase traces only a small fraction of a multiphase outflow whose larger mass and energy budget are carried by atomic and molecular gas \citep{crenshaw_feedback_2012, maskym_ufo_2023, winkel_close_2023, travascio_pds_2024}. Without radio--millimeter observations, however, this remains uncertain. Moreover, the commonly quoted 0.5--5\% benchmark assumes efficient conversion of injected AGN energy into outflow kinetic power, whereas coupling fractions of order 10--20\% would lower the required large-scale kinetic efficiency \citep{richings_cooling_2018}. Another possibility is temporal: the strong nuclear outflow and weak extended outflow may trace different AGN episodes, with a recent intense phase powering the compact wind and the larger-scale ionized gas reflecting an older, weaker phase. Even within one cycle, a recent high-accretion episode may not yet have built a stronger large-scale response. Using $R_{\rm out} \approx 0.3$\,kpc and $v_{\rm max} = 1150$ km s$^{-1}$, we estimate $t_{\rm dyn} = R/v_{\rm max} \approx 0.3$ Myr, consistent with an early feedback phase in which the strongest outflow signatures remain unresolved.
In this framework, Ton S180 is an instructive high-accretion Seyfert in which feedback is present but not yet efficient at galaxy-wide scales in the observed ionized phase. The resolved ionized outflow reaches the circumnuclear ring, so localized feedback in the central few kiloparsecs remains plausible, although higher angular resolution observations are needed to test this scenario. A luminous, rapidly accreting AGN phase therefore does not necessarily produce immediate kiloparsec-scale effects: feedback also depends on phase, geometry, and duty cycle as much as on the instantaneous accretion rate.

\section{Conclusions}
\label{sec:conclusions}
We present a detailed analysis of the local NLSy1 Ton S180 using VLT-MUSE integral field spectroscopy. By combining spectral modeling of the unresolved nucleus with a custom point-spread function subtraction algorithm, we isolated the faint emission of the host galaxy down to kiloparsec scales. This method relies on minimal assumptions and effectively handles the high contrast between the active nucleus and the host galaxy. Our main findings are summarized as follows.
\begin{enumerate}
  \item  The nuclear spectrum exhibits complex emission-line profiles that require a very broad component to model the permitted lines alongside a distinct outflowing component in the [\ion{O}{iii}] doublet. We derive black-hole masses in the range $\log(M_{\rm BH}/M_\odot)=6.5$--$7.7$ from the SE estimators retained in Table~\ref{tab:bhmass}. This range translates into markedly different dimensionless accretion rates, defined as $\dot{m} = \dot{M} / \dot{M}_{\rm{Edd}}$ (where $\dot{M}_{\rm Edd} \equiv L_{\rm Edd}/c^2$, without including the radiative efficiency factor, $\eta$), from an extreme $\dot{m} \sim 10^3$ solution at the low-mass end to $\dot{m} \sim 4$ at the high-mass end, but the source remains in a high-accretion regime in all cases.

  \item  Removing the strong nuclear emission reveals the structure and kinematics of the host galaxy. The H$\alpha$ emission traces a prominent circumnuclear ring connected to the spiral arms and an elongated inner structure compatible with a bar, while the large-scale gas and stellar kinematics remain rotation-dominated. The circular model leaves residuals up to $\sim 50$\,km\,s$^{-1}$; outside the ring, the residual pattern along the spiral arms is consistent with expectations for non-axisymmetric streaming motions directed toward the circumnuclear ring. Inside the ring, simple uniform or bar-driven inflow components neither improve the fit nor yield significant inflow velocities at the present spatial resolution. Therefore, we do not directly detect feeding toward the central engine, and higher-resolution IFU observations are needed to establish whether gas transport is at play on scales within the ring.
  
  \item  The spatially resolved emission line ratio diagnostics for gas ionization show that the spiral arms and circumnuclear ring mostly reflect a mix of stellar and AGN photoionization. In contrast, the region extending westward from the nucleus is clearly dominated by photoionization from the AGN and spatially coincides with the extended [\ion{O}{iii}] emission.

  \item We detect a spatially resolved ionized outflow propagating along this western structure up to two kiloparsecs from the nucleus. The unresolved nuclear outflow is strong, with $v_{\rm max} \sim 1150$\,km\,s$^{-1}$ and $\dot{M}_{\rm out} > 1.5\,M_\odot\,{\rm yr}^{-1}$ for the conservative wing-only estimate. Assigning the full blueshifted [\ion{O}{iii}] profile to the outflow instead gives $\dot{M}_{\rm out} > 3.2\,M_\odot\,{\rm yr}^{-1}$. The extended ionized component is remarkably faint, with $v_{\rm max} \sim 340$\,km\,s$^{-1}$ and a mass outflow rate of only $\sim 4.2 \times 10^{-4}\,M_\odot\,{\rm yr}^{-1}$. The ionized outflow kinetic coupling efficiency drops strongly from the unresolved nuclear to kiloparsec scales, implying that the optical phase alone carries only a modest fraction of the available outflow power.

  \item Taken together, these results indicate that Ton S180 is undergoing a rapid black hole growth episode in which the ionized outflow is detected but remains weak on host-galaxy scales. The present data favor a picture in which the observed optical outflow dissipates efficiently in the inner few kiloparsecs, couples only weakly to the large-scale interstellar medium, or traces only one phase of a multiphase outflow. The contrast between the strong nuclear outflow and the weak extended component may also indicate different episodes of AGN activity over time, with the compact outflow linked to a recent intense phase and the larger-scale ionized gas to an older, weaker episode. An additional possibility is that the source is being observed during an early stage of the current cycle, before a stronger large-scale response has developed.
\end{enumerate}

\begin{acknowledgements}
  PC, EC, GV, AT, and MB acknowledge the support of the ESO Scientific Visitor Programme.
  GV and SC acknowledge support from European Union's HE ERC Starting Grant No. 101040227 - WINGS.
  AT acknowledges financial support from the Bando Ricerca Fondamentale INAF 2022 Large Grant `Toward a holistic view of the Titans: multi-band observations of $z>6$ QSOs powered by greedy supermassive black holes' and from the Bando Ricerca Fondamentale INAF 2024 Large Grant `The DEepest study of LUminous QSOs in X-ray at z=2-7'.
  CR acknowledges support from SNSF Consolidator grant F01$-$13252, Fondecyt Regular grant 1230345, ANID BASAL project FB210003 and the China-Chile joint research fund.
  ET acknowledges support from ANID through CATA-BASAL FB210003, and FONDECYT Regular 1241005 and 1250821.
  AI tools (Gemini, Claude) were employed for language refinement and technical code assistance. The integrity of the research, including all analysis and results, was ensured through independent verification by the authors.
\end{acknowledgements}

\bibliographystyle{aa}
\bibliography{aa60868-26}

\begin{appendix}
\nolinenumbers

\section{The \texttt{prism} framework}
\label{appendix:prism}
The spectral analysis presented in this work was carried out with our in-house \texttt{prism} Python framework. The package is designed around \texttt{astropy}-compatible models and fitting workflows \citep{astropy_2022}, with optional access to robust optimization tools also available through \texttt{Sherpa} \citep{siemiginowska_sherpa_2024}. The \texttt{prism} repository can be found at \url{https://github.com/pier-astro/prism-spec}.
A first core functionality of \texttt{prism} is the treatment of the instrumental response. Each model tested in this work was corrected for instrumental broadening using a flux-conserving response formalism. For MUSE, the line spread function is well approximated by a Gaussian, with approximately $\sigma \sim 50$ km s$^{-1}$ at H$\alpha$ and increasing to about $\sigma \sim 65$ km s$^{-1}$ in the [\ion{O}{iii}]--H$\beta$ region (\href{https://www.eso.org/sci/facilities/paranal/instruments/muse/doc/ESO-261650_MUSE_User_Manual_P116.pdf}{MUSE User Manual}). In \texttt{prism}, this broadening is represented through a response matrix acting on the intrinsic model spectrum. This formulation redistributes the line flux over wavelength while preserving the integrated flux, allowing us to recover intrinsic widths without biasing the measured line luminosities. The same response formalism can be built, stored, and reapplied for different instruments and spectral samplings.
A second core functionality is the line-model framework itself. \texttt{prism} provides Gaussian, Lorentzian, and Voigt line components together with grouped line models that can be initialized directly from CSV line lists. This makes it straightforward to define physically motivated systems in which multiple transitions share velocity offsets and broadenings, while selected multiplet ratios are fixed to their theoretical or empirical values. This extends the flexible tied-line strategy introduced by \citet{ilic_fantastic_2023}. In our analysis, this was used both for the narrow forbidden-line systems and for the more complex broad-line decomposition. The \ion{Fe}{ii} calculations and multiplet prescriptions adopted in \texttt{prism} follow the treatment presented by \citet{ilic_fantastic_2023}.
Finally, \texttt{prism} was developed to handle both single spectra and full data cubes within the same environment. The same model definitions adopted for the unresolved nuclear spectrum can be propagated to spatially resolved fitting, with parallelized execution, robust optimizers, and structured outputs for parameter maps and fit diagnostics. The empirical AGN subtraction described in Appendix~\ref{appendix:psfmod} was implemented within the same framework, which allowed the subtraction products and the subsequent emission-line modeling to remain fully consistent throughout the analysis in a simple and user-friendly way.

\section{PSF modeling and AGN subtraction}
\label{appendix:psfmod}
We adopted a custom empirical subtraction strategy inspired by \citet{Marasco2020, travascio_pds_2024}, using the observed unresolved AGN spectrum as a spaxel-by-spaxel template and allowing for a smooth wavelength dependence in the PSF scaling. The active nucleus dominates the central emission of Ton S180, so this approach minimizes assumptions while remaining well matched to the contrast of the data. In the spectral windows used for the host-galaxy analysis (i.e., around the H$\beta$--[\ion{O}{iii}] and H$\alpha$--[\ion{N}{ii}] complexes), the effective PSF FWHM is about $0.6''$, consistent with the spatial resolution quoted in Sect.~\ref{sec:spatial_analysis}.
The reference nuclear spectrum was extracted from the four brightest spaxels around the emission peak. We preferred the observed spectrum itself over its parametric fit because the data-to-model mismatch, although below 4\%, is already comparable to the weak extended signatures of interest. In addition, the extremely high S/N of the nuclear spectrum ($\mathrm{(S/N)}_{\mathrm{pix}}(5100\,\text{\AA}) > 340$) ensures that the template noise is negligible relative to the noise in individual spaxels.
For each spaxel, we modeled the spectrum as the reference nuclear spectrum modulated by a third-degree B-spline, plus a third-degree polynomial accounting for the host-galaxy continuum. The polynomial degree was chosen as the best compromise between smoothness and flexibility in the fit spectral ranges. The B-spline modulation captures the wavelength-dependent PSF variations while remaining too smooth to mimic narrow emission features.
The full model is
\begin{equation}
  \label{eq:subtraction}
    S_{\rm{spax}}(\lambda) = S_{\rm{nucl}}(\lambda) \times \text{B-Spline}_3(\lambda) + P_3(\lambda),
\end{equation}
where $S_{\rm{spax}}$ is the model spectrum of a given spaxel, $S_{\rm{nucl}}$ is the reference nuclear spectrum, $\text{B-Spline}_3$ is the third-degree B-spline modulation, and $P_3$ is the third-degree polynomial continuum term. We used seven spline coefficients, corresponding to knot spacings of about 100\,\AA, which provided sufficient flexibility without introducing artificial narrow residuals (minimum FWHM reproduced by the B-spline $>$5000 km s$^{-1}$).
The subtraction was run separately in the H$\beta$--[\ion{O}{iii}] region (4610--5270\,\si{\angstrom}) and the H$\alpha$ region (6215--6875\,\si{\angstrom}).
Model fitting was performed spaxel by spaxel with a robust least-squares optimizer and iterative sigma-clipping to prevent emission lines and artifacts from biasing the continuum estimate. The procedure removes the unresolved AGN contribution together with the local continuum, which means that stellar-population information is intentionally lost in the inner region. For the present work this trade-off is acceptable, because the nuclear continuum does not show clear stellar absorption features and a full AGN--host decomposition would be highly degenerate.
\begin{figure}
  \centering
  \includegraphics[width=\linewidth]{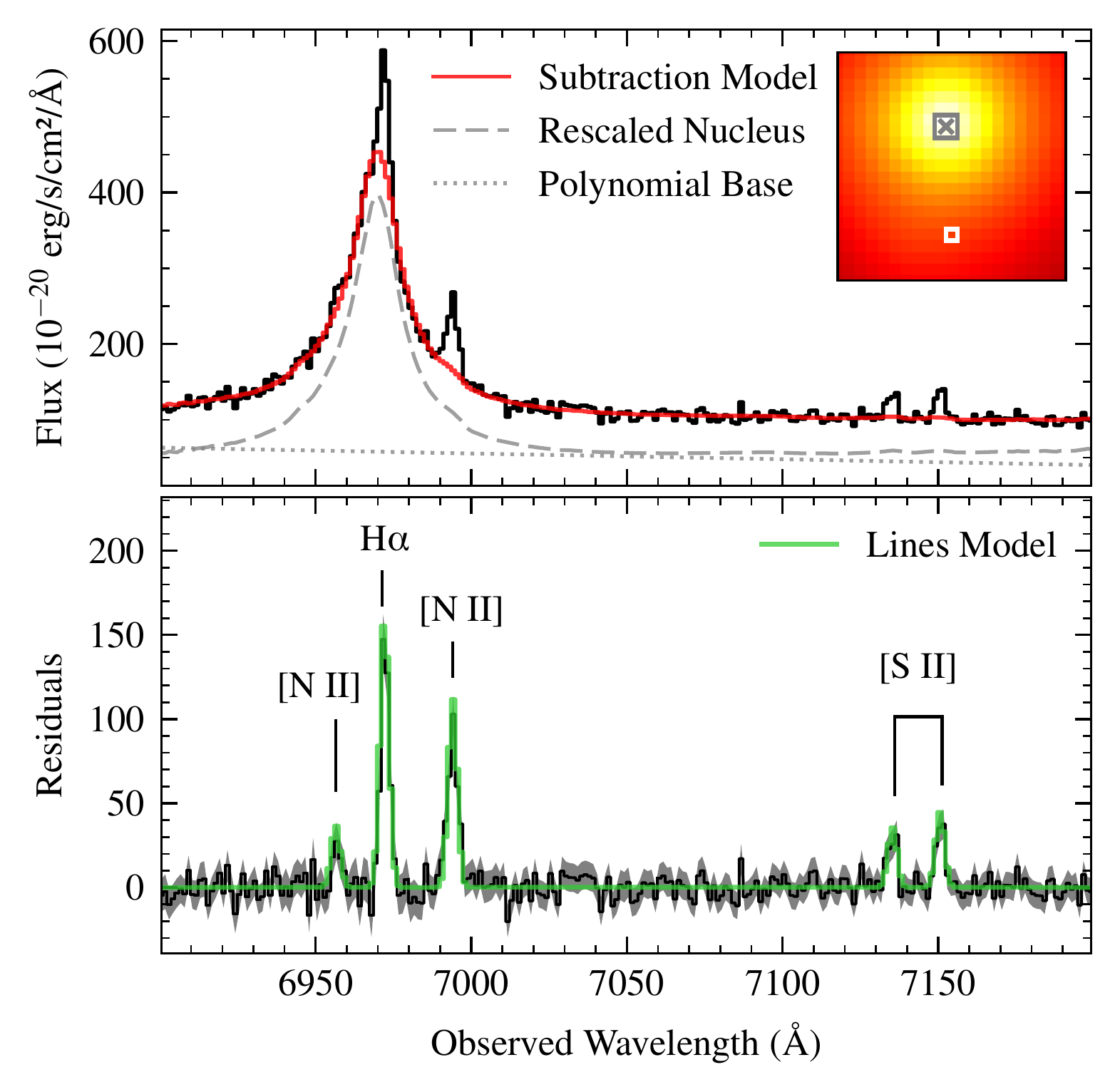}
  \caption{AGN PSF subtraction example for a spaxel located $\sim2$ kpc from the nucleus. \textit{Inset}: Zoomed-in image of the galaxy with the reference nuclear region highlighted as a gray cross box and the spaxel of interest marked with a white rectangle. \textit{Top}: Observed H$\alpha$ complex dominated by the AGN PSF wing in black with the nucleus subtraction model in red. The dashed line shows the reference nuclear spectrum scaled by the B-spline modulation to match the PSF contribution in this spaxel ($S_{\rm{nucl}}(\lambda) \times \text{B-Spline}_3$ in Eq.~\ref{eq:subtraction}). The dotted line shows the smooth polynomial component accounting for the host-galaxy continuum ($P_3(\lambda)$ in Eq.~\ref{eq:subtraction}). \textit{Bottom}: Residual spectrum with its error after subtraction, revealing the underlying galaxy emission lines, which are then fit with Gaussian profiles (green).}
  \label{fig:subtr_ex}
\end{figure}
We limited the analysis to a 100$\times$100 spaxel sub-cube centered on the nucleus and constructed a corresponding error cube that propagates both the spaxel variance and the uncertainty in the nuclear template scaling. For each spaxel, the variance spectrum is
\begin{equation*}
    \sigma_{\text{sub}}^2(\lambda) = \sigma_{\text{spax}}^2(\lambda) + \left[\sigma_{\text{nucl}}(\lambda) \cdot \text{B-Spline}_3(\lambda)\right]^2,
\end{equation*}
where $\sigma^2_{\text{spax}}$ is the variance of the spaxel spectrum and $\sigma^2_{\text{nucl}}$ is the variance of the reference nuclear spectrum. The polynomial term was neglected in the propagated variance because it acts mainly as a smooth flux offset.
As a consistency check, we measured the [\ion{N}{ii}] $\lambda6583/[\ion{N}{ii}]\lambda6548$ and [\ion{O}{iii}] $\lambda5007/[\ion{O}{iii}]\lambda4959$ ratios throughout the subtracted cube. Within the $1\sigma$ uncertainties, both remain consistent with the theoretical value of about three, supporting the reliability of the subtraction.

Before adopting this empirical approach, we tested a more conventional PSF reconstruction based on a field star template (marked with a green circle in Fig.~\ref{fig:CubeImage}). We modeled the stellar PSF with a 2D elliptical Moffat profile as a function of wavelength, fitting 20 integrated images between 5000 \AA\ and 8000 \AA\ while excluding the laser-affected region ($\lambda_{\rm{laser}0} = [5800, \, 5900]$~\AA). The best-fit parameters were then described with third-degree polynomials and used to reconstruct a wavelength-dependent PSF cube on the MUSE sampling, adapting routines from the Muse Python Data Analysis Framework (\texttt{MPDAF}, \citealt{MPDAF2017}). The method performs well on the template star itself, with a median subtraction error of about 3\%. However, when applied to Ton S180, it over-subtracts the broad H$\alpha$ and H$\beta$ emission, producing artificial absorption troughs and a coherent non-radial residual pattern along the putative bar. This failure may reflect either significant PSF variation across the MUSE field of view or a non-negligible, nearly featureless host contribution in the innermost spaxels. Tests on an independent MUSE dataset, where the AGN--host decomposition is better anchored by detectable host spectral features, indicate that the same strategy can perform satisfactorily in less extreme contrast conditions. For the present data, the empirical subtraction therefore provides the more robust solution.

\section{AGN spectral modeling and implications}
\label{appendix:agnmodel}

\begin{table}
\tiny
\centering
\caption{Best-fit parameters for the prominent emission lines in the unresolved AGN spectrum of Ton S180 for the Gaussian+Lorentzian (GL) model.}
\label{tab:AGNfit_GL}
\setlength{\tabcolsep}{4pt}
\renewcommand{\arraystretch}{1.2}
\begin{tabular}{lccc}

\toprule
Line & Flux & FWHM & Velocity Offset \\
\midrule
\multicolumn{4}{c}{Lorentzian BLR} \\
H$\beta$ 4861 & $158.1_{-1.9}^{+0.9}$ & $1501_{-6}^{+2}$ & $-203_{-7}^{+4}$ \\
H$\alpha$ 6563 & $561 \pm 9$ &  &  \\
\ion{He}{i} 5877 & $23.8 \pm 1.0$ & $2170_{-30}^{+10}$ & $-303_{-25}^{+8}$ \\
\ion{He}{ii} $\lambda$4686 Complex & $30 \pm 10$ & $>1500$ & $>100$ \\

\midrule
\multicolumn{4}{c}{narrow broad line region (nBLR)} \\
H$\beta$ 4861 & $30.0_{-0.9}^{+0.5}$ & $599_{-3}^{+6}$ & $-11_{-3}^{+7}$ \\
H$\alpha$ 6563 & $181_{-2}^{+5}$ &  &  \\
\ion{He}{i} 4471 & $1.8 \pm 0.2$ &  &  \\
\ion{He}{i} 5877 & $5.0 \pm 0.2$ &  &  \\
\ion{He}{i} 6678 & $1.42_{-0.06}^{+0.22}$ &  &  \\

\midrule
\multicolumn{4}{c}{Narrow Line Region (NLR)} \\
\ion{Na}{i} 5890 & $1.6 \pm 0.2$ & $290_{-30}^{+50}$ & $0 \pm 12$ \\
\ion{Na}{i} 5896 & $2.6_{-0.7}^{+0.2}$ &  &  \\
{[\ion{O}{i}]} 6300 & $0.7 \pm 0.1$ &  &  \\
{[\ion{N}{ii}]} 6583 & $21.2_{-0.7}^{+2.0}$ &  &  \\
{[\ion{S}{ii}]} 6716 & $1.6_{-0.3}^{+0.5}$ &  &  \\
{[\ion{S}{ii}]} 6731 & $1.2_{-0.3}^{+0.4}$ &  &  \\
{[\ion{O}{iii}]} 5007 (core) & $22.5_{-0.7}^{+0.3}$ & $363_{-4}^{+1}$ & $-135.1_{-2.7}^{+0.8}$ \\
{[\ion{O}{iii}]} 5007 (wing) & $17.4_{-0.7}^{+0.4}$ & $678 \pm 6$ & $-487_{-3}^{+6}$ \\

\midrule
\multicolumn{4}{c}{Lorentzian Iron Multiplets} \\
\ion{Fe}{ii} & --- & $1140_{-20}^{+10}$ & $-38_{-2}^{+7}$ \\
\bottomrule
\end{tabular}
\tablefoot{Fluxes are given in units of $10^{-15}$ erg s$^{-1}$ cm$^{-2}$; FWHM and velocity offsets are given in kilometers per second. Errors are at the 68\% confidence level.}
\end{table}

\begin{figure} 
  \centering
  \includegraphics[width=\linewidth]{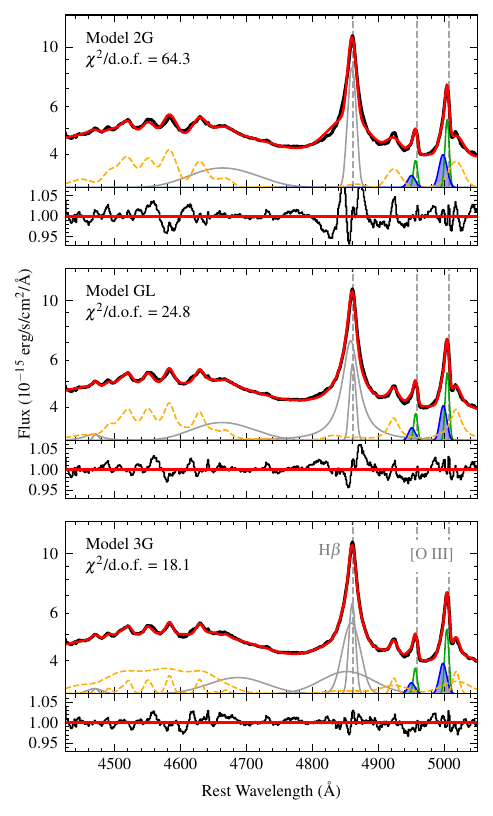}
  \caption{Comparison of the three best-fit spectral models for the nuclear spectrum with a zoom-in in the H$\beta$ line region. The fiducial three-Gaussian (3G) model (\textit{bottom}) best reproduces the observed profile, capturing both the line core and wings. The Gaussian+Lorentzian (GL) model (\textit{middle}) provides a reasonable fit but misses some details. The two-Gaussian (2G) model (\textit{top}) provides the worst fit. Each panel shows the data (black), total model (red), and individual components, with gray for Balmer, green for [O~\textsc{iii}], blue for the [O~\textsc{iii}] blue wing, and dashed yellow for the \ion{Fe}{ii} model components.  The data-to-model ratios are shown below each fit. Although not visible, errors are plotted together with the data. The top left corner of each panel indicates the reduced chi-squared value for the fit. The vertical dashed gray lines mark the wavelengths of H$\beta$ and [O~\textsc{iii}] at the systemic velocity.}
  \label{fig:modelcomp}
\end{figure}
Here we summarize the alternative fits and line-profile tests that support the fiducial nuclear model adopted in Sect.~\ref{sec:agnspec}, and we discuss their implications for the black hole mass estimates.
We tested several models of the unresolved nuclear spectrum over both the broad band (4425--7125~\si{\angstrom}) and the H$\beta$-focused window (4480--5100~\si{\angstrom}). The broad band provides the most physically complete description of the spectrum, while the narrower window isolates the Balmer--\ion{Fe}{ii} complex and is useful to test the robustness of the line-profile decomposition. The best-fit parameters are consistent across the two fitting ranges, although the reduced $\chi^2$ is about twice as high in the broad band because of residual continuum mismatches and weak secondary features outside the main emission complexes.
The main alternative models are compared in Fig.~\ref{fig:modelcomp}. A two-Gaussian model (2G), with only narrow and broad Balmer components and a single Gaussian for \ion{Fe}{ii}, provides a poor fit ($\chi^2/\mathrm{d.o.f.} \approx 68.2$ with 59 free parameters). A Gaussian+Lorentzian model (GL), in which the vBLR and iBLR components are replaced by a single Lorentzian for the Balmer, He, and \ion{Fe}{ii} lines, performs substantially better ($\chi^2/\mathrm{d.o.f.} \approx 24.8$ with 64 free parameters; Table~\ref{tab:AGNfit_GL}) but still misses part of the observed line structure. The fiducial three-Gaussian model (3G), with vBLR, iBLR, and nBLR components for the permitted lines and two Gaussians for \ion{Fe}{ii}, gives the best description ($\chi^2/\mathrm{d.o.f.} \approx 18.1$ with 81 free parameters), with relative residuals within 4\% and RMS $2.8 \times 10^{-18}$ erg s$^{-1}$ cm$^{-2}$ \si{\angstrom}$^{-1}$.
Although NLSy1 permitted lines are often described with Lorentzian profiles \citep{berton_line_2020}, the BIC strongly favors the 3G solution over the GL model, with $\Delta$BIC = BIC$_{\rm GL}$ -- BIC$_{\rm 3G} > 500$. Even if the spectral errors are doubled to account conservatively for possible underestimation, the preference for the 3G model remains well above the threshold for strong evidence. We also tested broken power laws and exponential wings for the broad Balmer emission, but neither improved the fit. Additional \ion{Fe}{ii} flexibility yields only modest gains: introducing a third \ion{Fe}{ii} component improves the H$\beta$--\ion{Fe}{ii} fit from $\chi^2/\mathrm{d.o.f.} = 10.8$ to 9.7, whereas tying \ion{Fe}{ii} kinematics to the Balmer kinematics degrades it only marginally to 11.1. Because the extra \ion{Fe}{ii} component increases degeneracy in the blended iron complex, we retain the simpler and more reproducible two-component \ion{Fe}{ii} description.
The 4640--4700 \AA\ blend deserves separate comment. In the fiducial fit we model it as an unresolved complex that may include He II $\lambda$4686, the Bowen fluorescence lines N III $\lambda$4640 and C III $\lambda$4650 \citep[e.g.,][]{bowen_fluorescence_2023}, and underlying Fe II emission broadened by the BLR kinematics. When the individual amplitudes are left free, the decomposition becomes highly degenerate. Forcing the feature to be represented mainly by a single He II component drives the fit toward unphysical solutions in which the inferred He II flux becomes comparable to or larger than that of H$\beta$. We therefore report only the integrated flux of the complex in Table~\ref{tab:AGNfit}.
These profile choices propagate directly into the black hole mass estimates discussed in Sect.~\ref{sec:results}. In all virial-mass calculations we include the narrower Balmer component because its FWHM ($\sim 610$ km s$^{-1}$) is substantially larger than that of the narrow forbidden lines, suggesting that it does not arise in the host-galaxy interstellar medium. Its inclusion is also supported by the detection of a \ion{Fe}{ii} component with similar width, consistent with an origin in the BLR. By contrast, a purely narrow Balmer component fixed to the kinematics of the forbidden lines is not statistically required and contributes negligibly to the total line flux.
Our MUSE data also reveal a very broad Balmer component with FWHM $> 4500$\,km\,s$^{-1}$ that was not isolated in previous lower-S/N spectra. A comparison with the H$\alpha$ and H$\beta$ profiles from \citet{scharwachter_spatially_2017} indicates that this component is not in conflict with their data; rather, it was hidden within the noise and lower spectral quality of the earlier WiFeS observations. Its inclusion is further supported by the kinematic alignment with the \ion{Fe}{ii} emission, and fits in which the \ion{Fe}{ii} kinematics are tied to the Balmer kinematics recover a similar broad component with FWHM $\sim 4200$\,km\,s$^{-1}$. Regardless of its physical interpretation, this vBLR component is required to reproduce the observed Balmer wings in a Gaussian decomposition.

The effect of the vBLR component on the SE mass estimates is detailed in Table~\ref{tab:se_mass_breakdown}, where we report the line luminosities and line widths (FWHM and velocity dispersion, $\sigma$, i.e., the moment 2) of the total BLR emission line profile, and derived black hole masses both with and without the vBLR component.
For the FWHM-based calibration of \citet{woo_new_2026} (Eq.~12), including the vBLR increases the mass estimate only slightly from $\log(M_{\rm BH}/M_\odot) = 6.4 \pm 0.4$ to $6.5 \pm 0.4$. In contrast, for the mean-spectrum line-dispersion calibration of \citet{woo_new_2026} (Eq.~14), including the vBLR raises the mass estimate from $\log(M_{\rm BH}/M_\odot) = 6.7 \pm 0.4$ to $7.7 \pm 0.4$, resulting in a large discrepancy of $\sim 1.26$\,dex between the two estimates.

This substantial discrepancy arises from the intrinsic difference in sensitivity between FWHM and $\sigma$ to broad wings in non-Gaussian line profiles. The FWHM is defined at half the peak intensity, which is set primarily by the core of the line profile (the iBLR+nBLR blend in our case). Adding the underlying vBLR base increases the overall peak height slightly and does not significantly affect the shape of the line core, resulting in a minor change in FWHM (from $872$ to $963$\,km\,s$^{-1}$, corresponding to $\Delta \log \text{FWHM} = 0.043$\,dex). Because Eq.~12 of \cite{woo_new_2026} scales as $\text{FWHM}^{2.68}$, this small shift translates into an increase of only $+0.12$\,dex in mass. Conversely, the line dispersion $\sigma$ is the square root of the second moment of the profile, which scales quadratically with velocity offset from the line centroid. The addition of the high-velocity vBLR wings increases $\sigma$ from $635$ to $1469$\,km\,s$^{-1}$ (a factor of $2.31$, or $\Delta \log \sigma = 0.364$\,dex). Because Eq.~14 scales as $\sigma^{2.89}$, this shift produces an increase of $+1.05$\,dex in the estimated mass.

\begin{table*}
\caption{Single-epoch black hole mass input line parameters and resulting mass estimates.}
\label{tab:se_mass_breakdown}
\centering
\small
\begin{tabular}{lcccccc cc}
\toprule
Line & Components & Luminosity & FWHM & $\sigma$ & \multicolumn{2}{c}{\citet{dalla_bonta_sloan_2020, dalla_bonta_estimating_2025}} & \multicolumn{2}{c}{\citet{woo_new_2026}} \\
\cmidrule(lr){6-7} \cmidrule(lr){8-9}
& & ($10^{42}\text{ erg s}^{-1}$) & ($\text{km s}^{-1}$) & ($\text{km s}^{-1}$) & $\log M_{\rm FWHM}$ & $\log M_\sigma$ & $\log M_{\rm FWHM}$ & $\log M_\sigma$ \\
\midrule
H$\beta$ & vBLR + iBLR + nBLR & $1.79 \pm 0.07$ & $960 \pm 50$ & $1470 \pm 50$ & $7.2 \pm 0.4$ & $7.7 \pm 0.3$ & $6.5 \pm 0.4$ & $7.7 \pm 0.4$ \\
H$\beta$ & iBLR + nBLR        & $1.17 \pm 0.06$ & $870 \pm 40$ & $640 \pm 30$  & $7.0 \pm 0.4$ & $6.7 \pm 0.3$ & $6.4 \pm 0.4$ & $6.7 \pm 0.4$ \\
\midrule
H$\alpha$ & vBLR + iBLR + nBLR & $7.1 \pm 0.3$ & $860 \pm 40$ & $1340 \pm 50$ & $7.1 \pm 0.3$ & $7.6 \pm 0.2$ & \multicolumn{2}{c}{---} \\
H$\alpha$ & iBLR + nBLR        & $5.1 \pm 0.3$ & $810 \pm 40$ & $610 \pm 30$  & $7.0 \pm 0.3$ & $6.6 \pm 0.2$ & \multicolumn{2}{c}{---} \\
\bottomrule
\end{tabular}
\tablefoot{Emission-line parameters and SE black hole mass estimates ($\log(M_{\rm BH}/M_\odot)$) derived with and without the very broad component (vBLR). Luminosities refer to the integrated emission-line flux corrected for Galactic extinction ($E(B-V)_{\rm Gal} = 0.0124$\,mag). The monochromatic continuum luminosity at 5100\,\AA\ used in the \citet{woo_new_2026} relations (Eqs.~12 and 14) is $L_{5100} = (2.03 \pm 0.05) \times 10^{44}\text{ erg s}^{-1}$. Virial factors adopted are $\langle \log f \rangle = 0.68 \pm 0.03$ for \citet{dalla_bonta_sloan_2020, dalla_bonta_estimating_2025}, $\log f = 0.05 \pm 0.12$ for the FWHM calibration in \citet{woo_new_2026}, and $\log f = 0.65 \pm 0.12$ for the $\sigma$ calibration in \citet{woo_new_2026}. The inversion $\sigma > \text{FWHM}$ upon including the vBLR occurs because line dispersion is weighted by $(\lambda - \lambda_0)^2$ and thus strongly driven by high-velocity wings, whereas FWHM is anchored to the narrow line core.}
\end{table*}

Furthermore, \citet{woo_new_2026} emphasize that their velocity-dispersion and FWHM calibrations are obtained from a sample having an average line-profile shape with a typical ratio $\text{FWHM}/\sigma_{\rm mean} = 2.0 \pm 0.6$. For Ton S180, excluding the vBLR component yields $\text{FWHM}/\sigma = 1.37$, leading to $M_{\rm BH}$ estimates consistent within $0.32$\,dex between Eqs.~12 and 14 of \citet{woo_new_2026}. However, when the vBLR component is included, the profile ratio drops to $\text{FWHM}/\sigma = 0.655$, deviating significantly from the empirical calibration sample of \citet{woo_new_2026} and producing the observed $\sim 1.26$\,dex offset. A similar behavior is observed for the \citet{dalla_bonta_sloan_2020, dalla_bonta_estimating_2025} relations, where excluding the vBLR component lowers the estimated masses by $0.2\text{--}0.97$\,dex depending on the line and width tracer and makes them more consistent among the adopted velocity-width measures (either FWHM or $\sigma$).

The older SE estimate of \citet{scharwachter_spatially_2017}, based on H$\beta$ FWHM without the vBLR component, gave $\log(M_{\rm BH}/M_\odot) = 6.88 \pm 0.05$. The recent X-ray reverberation estimate from the Fe~K$\alpha$ lag, $\log(M_{\rm BH}/M_\odot) = 7.46_{-0.35}^{+0.01}$ \citep{kumar_reverberation_2026}, lies near the upper end of our range. These comparisons underscore that, in the absence of direct optical reverberation mapping, SE mass estimates in super-Eddington AGNs are subject to strong systematic uncertainties governed by the line profile shape (see Sect.~\ref{sec:agnspec}), line-width definitions, and the choice of the calibrator.

\section{Electron density estimation}
\label{appendix:density}
Density estimates are crucial for deriving the physical properties of the outflow, yet they are often uncertain due to the limitations of available diagnostics. In our data, we have access to the [\ion{S}{ii}] $\lambda\lambda$6716, 6731 doublet, which is a common density diagnostic for ionized gas. However, this doublet traces relatively low-density gas and may not accurately reflect the conditions in the outflowing component.
We investigated the electron density using the [\ion{S}{ii}] doublet ratio \citep{osterbrock_astrophysics_2006} and the ionization parameter method proposed by \citet{baron_discovering_2019}.
We used as reference spectrum the one extracted from a circular aperture in the region where the outflow is detected in the AGN-subtracted cube (see Sect.~\ref{sec:outflow}). This spectrum is dominated by host-galaxy emission and the wind is detected only as a weak blue wing in [\ion{O}{iii}], but it is the best available proxy for the conditions in the outflow region given the lack of direct diagnostics for the outflowing gas.
Section \ref{sec:bpt} shows that this region is AGN-ionized, so the physical conditions are likely dominated by the AGN radiation field, making it a reasonable approximation for the outflowing gas.
The nuclear [\ion{S}{ii}] ratio ($\lambda6716/\lambda6731 \approx 1.34$) implies a density of $n_e \approx 80$\,cm$^{-3}$ assuming a temperature of $T_e = 10^4$\,K. However, this value is highly uncertain as the ratio approaches the low-density saturation limit and likely underestimates the density since [\ion{S}{ii}] traces more diffuse gas. In the extended outflow region, the [\ion{S}{ii}] doublet ratio yields $n_e \approx 255$\,cm$^{-3}$.
To obtain a more representative value for the AGN-driven wind conditions, we estimated $n_e$ from the ionization parameter $U$. The ionization parameter is derived from extinction-corrected emission line ratios using Eq. 2 of \citet{baron_discovering_2019},
\begin{equation}
  \log U = -3.77 + 0.19x + 0.78x^2 - 0.25y + 0.34y^2,
\end{equation}
where $x = \log(\rm [\ion{O}{iii}]/H\beta)$ and $y = \log(\rm [\ion{N}{ii}]/H\alpha)$.
For Ton S180, we find $\log U = -3.57$. The electron density is then given by \citep[Eq. 4]{baron_discovering_2019}
\begin{equation}
    n_e \simeq 3.2 \left( \frac{L_{\rm bol}}{10^{45}\,{\rm erg\,s}^{-1}} \right) \left( \frac{R_{\rm out}}{\rm kpc} \right)^{-2} \left( \frac{1}{U} \right) \, {\rm cm}^{-3}.
\end{equation}
We derive $n_e = 600 \pm 400$\,cm$^{-3}$. This value is higher than the [\ion{S}{ii}]-based estimate and consistent with the standard assumption of $500$\,cm$^{-3}$ we adopt for the outflow calculations, supporting the validity of our approach.

\end{appendix}

\end{document}